\documentclass[preprint,10pt]{elsarticle}

\usepackage{graphicx}
\usepackage{subcaption}
\usepackage{amssymb}
\usepackage{amsmath}
\usepackage{bm}
\biboptions{numbers,sort&compress}
\usepackage[utf8]{inputenc}
\usepackage[english]{babel}
\usepackage{amssymb}
\usepackage{natbib}
\usepackage{longtable}
\usepackage{geometry}
\usepackage{longtable}

\biboptions{sort&compress}
\journal{Nuclear Physics A}

\begin{document}

\begin{frontmatter}

\title{Correlation between the Shape Coexistence and Stability in Mo and Ru isotopes}

\author[a]{Mamta Aggarwal}
\author[b]{G. Saxena}
\author[a]{Pranali Parab}

\address[a]{Department of Physics, University of Mumbai, Vidhyanagari, Mumbai, 400098, Maharashtra, India.}
\address[b]{Department of Physics (H \& S), Govt. Women Engineering College, Ajmer, Rajasthan$-$305002, India.}

\begin{abstract}
  In a rapidly changing shape phase region, the presence of shape coexistence and its possible impact on the decay modes and half$-$lives, has been explored in astrophysically interesting Mo and Ru isotopes, in an extensive study within the microscopic theoretical framework using Nilsson Strutinsky Method and Relativistic Mean Field Model. The isotopic chains of Mo and Ru exhibit rapid shape phase transitions, triaxial $\gamma$ softness, shape instability along with many coexisting states mostly with oblate and triaxial shapes. Proton and neutron separation energies have been calculated and compared with the available data. Results obtained from both the formalisms are in good agreement with  each other as well as the available experimental data. Our computed  $\beta-$decay half$-$lives and separation energy for nuclei exhibiting shape coexistence were examined for the decay mode from second minima state of the parent nuclei to the ground or excited state of the daughter nuclei. The second minima state of the coexisting shapes in Mo and Ru isotopes, were seen to impact the structural properties, $\beta-$decay half$-$lives, separation energy and hence the stability of the nuclei.\par

\end{abstract}


\end{frontmatter}

\section{Introduction}
\label{sec:Introduction}
Precise measurements and the knowledge of various nuclear properties like  nuclear masses, decay mechanisms and the half$-$lives of many unstable nuclei lying  far from the stability valley are crucial inputs for the extensive investigations by the astrophysicists to understand the energy production in stars, supernova explosion and the nucleosynthesis r$-$process ~\cite{Bethe,Burbidge,Baade}. Most of the nuclear structural inputs that are relevant in the modeling of nucleosynthesis r$-$process path are still scarce experimentally and one needs to rely on the theoretical estimates which in turn provide a stringent test for the theoretical models of nuclear structure, trusted for their predictions in the stable regions.  The nuclei in mass region A $\approx$ 80 $-$ 120 especially the Zr, Mo and Ru isotopes characterized by the rapid structural changes and shape instabilities, are located at the key points in the r$-$process path ~\cite{Nabi} and play an important role in the astrophysical nuclear synthesization process ~\cite{Burbidge,Cowan,Sasaki}, which is why this region is of special interest to theoretical nuclear physicists to explore a variety of distinct shapes and structural transitions as well as to study their impact on stability, decay  modes and their life times.\par

An atomic nucleus, being a finite quantum many$-$body system, can display a variety of distinct geometrical shapes along with the coexisting shapes ~\cite{MAPRC89} that are known to provide good evidence for the crucial role of underlying shell structure and nuclear collectivity in the stability and  structure of nuclei that has been amply demonstrated in several studies of nuclear decay and ﬁssion processes ~\cite{Strut1967,Strut1968,Nilsson,MollerNix,Yoshida}. Various phenomena of nuclear coexistence like high$-$K isomers, superdeformed states, shape and pairing isomers, and low$-$lying deformed states, manifesting as close$-$lying nuclear eigenstates, exhibit very different intrinsic properties or shapes and have been attracting considerable attention in the recent times. Nuclear shape coexistence ~\cite{Werner,Simpson,Nomura}, in particular, now known to exist in various domains of the nuclear chart, is characterized by one or more states lying close to the ground state having different deformations with the competing spherical, axially symmetric prolate and oblate, and triaxial shapes ~\cite{Ring} at very close or similar energies. A nucleus with coexisting eigenstates, exists in a second minima state lying close to the ground state that may have lifetime sufficient for the probes. The lifetime of the second minima state will depend on the extent of overlap between its wave function with that of the ground state, its excitation energy and the height of saddle separating it from the ground state. A long lifetime may give rise to a meta$-$stable state or even a shape isomer. Such possible influence of the coexisting shapes on the lifetimes ~\cite{Crider}, as shown for the case of 2p$-$radioactivity in our recent work ~\cite{GMAJJPGlife}, invokes the present theoretical investigation that may provide useful inputs for probes as well as for the astrophysical studies. \par

Neutron$-$rich isotopes of Sr, Zr, Mo and Ru in and around the mass region A $\approx$ 80 $-$ 120 ~\cite{Hua,Sarriguren,Abusara} present an ideal and unique laboratory to study the coexistence of different shapes in a single atomic nucleus, the phenomena of rapid shape phase transitions, shape instabilities and shape mixing as a function of nucleon number, excitation energy and the angular momentum ~\cite{MAPRB693, MAIJMP28}. Among the several islands of axial symmetry in shapes on the nuclear chart discovered so far  ~\cite{MollerBengtsson}, one finds the domains of triaxial deformations ~\cite{Xiang} exhibiting the softness of nuclear potential toward triaxiallity in the isotopic and isotonic chains ~\cite{Cejnar} of these nuclei. These diverse features of nuclear shapes in this mass region, as also indicated by spectroscopic measurements ~\cite{Sumikama,Goodin,Urban,Campbell,Charlwood}, reflect the rearrangement of nucleons ~\cite{Heyde,Butler,Sorlin,Cejnar,Bender} near the fermi level arising as a consequence of the interplay between the single particle energies and collective degrees of freedom. Althgough it is not easy to trace the triaxial deformation experimentally, but some efforts like measurements of high spin states ~\cite{Marshalek,Odegard} and chiral rotations ~\cite{Frauendorf,Grodner,Meng,Ayangeakaa} have indicated the existence of triaxiallity ~\cite{Stachel,Zamfir,Sheikh}. Breaking of axial symmetry in the nuclear intrinsic states seen in this mass region (A $\approx$ 80 $-$ 100), influences the dynamical and structural properties  ~\cite{Yao,Fu} like  increasing the binding of nucleons ~\cite{MollerBengtsson} and lowering of the fission barrier of heavy nuclei ~\cite{Lu}. This points towards the importance of shape evolution including triaxiallity and the shape coexistence in nuclear decay properties of astrophysically relevant nuclei belonging to this mass region, which needs attention and a detailed investigation that frames the objective of the present work. \par

Here, we first explore the shape evolution and the phenomenon of shape coexistence in Mo and Ru  ~\cite{Nabi} isotopes that have a special significance in astrophysical processes and our understanding of solar system formation ~\cite{Dauphas}. These nuclei also have been a focus of nuclear structure research ~\cite{Nabi,Hukkanen,Karmakar,Abusara} in the recent times for the variety of shapes and structural transitions they are known to exhibit. We employ a well known simple yet effective theoretical microscopic approach using Nilson Strunsiky method (NSM) ~\cite{MAPLB} to perform a systematic study to explore shapes and shape coexistence in the entire isotopic chain $^{78-144}$Mo and $^{84-152}$Ru ranging from the neutron deficient to neutron rich isotopes lying in the ground state. Effects of temperature and spin on rapid shape phase transitions and shape coexistence shown in our earlier works for rare earth nuclei ~\cite{MAPRC90}, are not included here and will be discussed in our subsequent communications.  In addition, we also perform mean$-$field calculations employing an another well established formalism using Relativistic Mean$-$Field (RMF) theory \cite{saxenaPLB2019,saxena2017,saxenaplb2017} with NL3$^{*}$ parameter ~\cite{lala} which is widely used and found  to provide an excellent match with the experimental data in the considered mass region~\cite{bhuyan2015,Abusara,shi2018}. To further validate our results and check the parameter dependency of our RMF results, we use another variant of RMF i.e. density$-$dependent point coupling interaction (DD$-$PC) ~\cite{niksic08} for a few selected nuclei, which show a reasonably good match. This study of Mo, Ru isotopic chain displays the rapid shape phase transitions, triaxial $\gamma$ softness, shape instability and shape coexistence  as predicted by both our formalisms (NSM and RMF theory). Now, since the  $\beta-$decay half$-$life is sensitive to the shell structure near the Fermi level and is an indicator of the shape phase transition ~\cite{Yoshida}, we compute $\beta-$decay half$-$lives for nuclei exhibiting shape coexistence, and examine their variation due to decay from the second minima excited states to the ground state or excited state of the daughter nuclei. The possible impact of the second coexisting state lying close to the ground state on our evaluated $\beta-$ decay half$-$lives and also on the proton and neutron  separation energies has been studied. The results have been compared with the existing data and the predictions of various other theoretical models. The theoretical formalism (NSM) employed here being one of the most well established model known for a few decades now, again proves to be very effective in predicting the shapes and coexisting shapes with the emerging and competing $\gamma$ degree of freedom and triaxial softness  well displayed in this work. Moreover, the results obtained with the relativistic mean$-$field theory (RMF) using the parameters NL3* and DD$-$PC are in good match with those from NSM and provide us a well needed impetus to extend our calculations to more unknown territories of the periodic chart. RMF calculations are particularly important when dealing with nuclei in the mid$-$shell region because RMF model assists in understanding the behaviors of these nuclei under extreme conditions encountered in astrophysical environment. Also, a good match with the results of NSM and both the parameters of RMF show model independency and reliability of our results. The impact of shape instabilities seen on the stability of the nucleus is an important outcome of this work. \par

\section{Theoretical Formalism}
\label{sec:Theoretical Formalism}
\subsection{Nilsson Strunsiky Method}
Shape of an atomic nucleus is governed by the delicate interplay of macroscopic bulk properties of nuclear matter and the microscopic shell effects for which we use triaxially deformed Nilsson model along with the Strutinsky's prescription which starts with the Strutinsky density distribution function ~\cite{Strut1967,Strut1968,Brack,Nilsson} for single particle states as
\begin{equation}
\tilde {g(\epsilon)} = {1 \over {\sqrt{\pi}\gamma}} \sum exp(-u_i)^2 \sum_{k=0}^{\infty} C_k H_k (u_i),
\end{equation}
where
\begin{equation}
u_i = {{(\epsilon - \epsilon_i)} \over \gamma},
\end{equation}
and the coefficients C$_k$ are
\begin{equation}
    C_{k} = \begin {cases}
    \frac{(-1)^{\frac{k}{2}}}{2^{k}\frac{k}{2}!}, \hspace{0.2cm} \forall \hspace{0.1cm} even \hspace{0.1cm} k;\\
   0, \hspace{0.9cm} \forall \hspace{0.1cm} odd \hspace{0.1cm} k.
    \end{cases}
\end{equation}
Hermite polynomials H$_k$(u$_i$) upto higher order of correction are used to ensure better smoothening of the levels. The energy due to Strutinsky's smoothened single particle level distribution is given by
\begin{equation}
\tilde E = \int_{-\infty}^{\mu} \tilde g(\epsilon) d\epsilon
\end{equation}
The chemical potential $\mu$ is fixed by the number conserving equation
\begin{equation}
N = \int_{-\infty}^{\mu} \tilde g(\epsilon) d\epsilon
\end{equation}
The shell correction to the energy is obtained as
\begin{equation}
\delta E = \sum_{i=1} ^A \epsilon_i- \tilde E
\end{equation}
The first term is the shell model energy in the ground state and the second term is the smoothed energy with the smearing width 1.2$\hbar$ $\omega$. The single particle energies $\epsilon_i$ as a function of  deformation parameters ($\beta_{2}$, $\gamma$) are generated by Nilsson Hamiltonian for the deformed oscillator diagonalized in a cylindrical representation ~\cite{GsN,GsL,Eis}.
\begin{eqnarray}
H=p^2 /2m + (m/2)(\omega_x^2 x^2 + \omega_y^2 y^2 +\omega_z^2 z^2)+  \nonumber\\
 C l.s + D (l^2- 2<l^2>)
\end{eqnarray}
The coefficients for the $l.s $ and $l{^2}$ terms are taken from Seeger~\cite{Seg} who has fitted them to reproduce the shell corrections~\cite{Strut1967,Strut1968} to ground state masses. Strutinsky's  shell correction to energy  $\delta$E added to macroscopic energy of the spherical drop $BE_{LDM}$ reproduced by LDM  mass formula~\cite{Pm} along with the deformation energy $E_{def}$ (obtained from surface and coulomb effects) gives the total energy  $BE_{gs}$ corrected for microscopic effects of the nuclear system
\begin{eqnarray}
BE_{gs}(Z,N,\beta_{2},\gamma) = BE_{LDM}(Z,N) -  \nonumber\\
 E_{def}(Z,N,\beta_{2},\gamma)-\delta E_{shell}(Z,N,\beta_{2},\gamma)
\end{eqnarray}
To trace the equillibrium deformation and nuclear shapes, the energy $E (= -BE)$ minima are searched for various $\beta_{2}$ and $\gamma$  with axial deformation parameter $\beta_{2}$ varying from 0 to 0.4 in steps of 0.01 and the angular deformation parameter $\gamma$ ranging from $-180^o$ (oblate) to $-120^o$ (prolate) and triaxial in between. The precise position of the first unbound proton and neutron is located by one proton/neutron separation energy approaching zero value which we obtain by computing the difference between the binding energies $BE_{gs}$ of the parent and daughter nucleus.\par

\subsection{Relativistic Mean$-$Field (RMF) model}
Two classes of mean$-$field models are employed for this investigation:- the nonlinear meson$-$ nucleon coupling model (NL), and the density$-$dependent point$-$coupling (DD$-$PC) model. The main differences between these models consist in the treatment of the range of the interaction and in the density dependence. The density dependence is introduced via non$-$linear meson couplings (NL \cite{Lalazissis09}) or either through an explicit dependence of the coupling constants (DD$-$ME \cite{Lalazissis05} \& DD$-$PC \cite{niksic08}). At present, many major classes of covariant energy density functionals exist dependent on the combination of above mentioned features (see Ref. \cite{agbemava2014} for detail).\par
Mainly our RMF calculations have been carried out using the model Lagrangian density with nonlinear terms both for the ${\sigma}$ and ${\omega}$ mesons as described in detail in Refs. \cite{Lalazissis09,Boguta77,Boguta83,Furnstahl97}.
\begin{eqnarray}
       {\cal L}& = &{\bar\psi} [\imath \gamma^{\mu}\partial_{\mu}
                  - M]\psi\nonumber\\
                  &&+ \frac{1}{2}\, \partial_{\mu}\sigma\partial^{\mu}\sigma
                - \frac{1}{2}m_{\sigma}^{2}\sigma^2- \frac{1}{3}g_{2}\sigma
                 ^{3} - \frac{1}{4}g_{3}\sigma^{4} -g_{\sigma}
                {\bar\psi}  \sigma  \psi\nonumber\\
               &&-\frac{1}{4}H_{\mu \nu}H^{\mu \nu} + \frac{1}{2}m_{\omega}
                  ^{2}\omega_{\mu}\omega^{\mu} + \frac{1}{4} c_{3}
                 (\omega_{\mu} \omega^{\mu})^{2}
                  - g_{\omega}{\bar\psi} \gamma^{\mu}\psi
                 \omega_{\mu}\nonumber\\
              &&-\frac{1}{4}G_{\mu \nu}^{a}G^{a\mu \nu}
                 + \frac{1}{2}m_{\rho}
                 ^{2}\rho_{\mu}^{a}\rho^{a\mu}
                  - g_{\rho}{\bar\psi} \gamma_{\mu}\tau^{a}\psi
                 \rho^{\mu a}\nonumber\nonumber\\
               &&-\frac{1}{4}F_{\mu \nu}F^{\mu \nu}
                 - e{\bar\psi} \gamma_{\mu} \frac{(1-\tau_{3})}
                 {2} A^{\mu} \psi\,\,,
\end{eqnarray}
where the field tensors $H$, $G$ and $F$ for the vector fields are
defined by
\begin{eqnarray}
                 H_{\mu \nu} &=& \partial_{\mu} \omega_{\nu} -
                       \partial_{\nu} \omega_{\mu}\nonumber\\
                 G_{\mu \nu}^{a} &=& \partial_{\mu} \rho_{\nu}^{a} -In t
                       \partial_{\nu} \rho_{\mu}^{a}
                     -2 g_{\rho}\,\epsilon^{abc} \rho_{\mu}^{b}
                    \rho_{\nu}^{c} \nonumber\\
                  F_{\mu \nu} &=& \partial_{\mu} A_{\nu} -
                       \partial_{\nu} A_{\mu}\,\,\nonumber\
\end{eqnarray}
and other symbols have their usual meaning. The corresponding Dirac equations for nucleons and Klein$-$Gordon equations for mesons obtained with the mean$-$field approximation are solved by the expansion method on the widely used axially deformed Harmonic$-$Oscillator basis \cite{Geng2003,Gambhir1989}. The quadrupole constrained calculations have been performed for all the nuclei considered here in order to obtain their potential energy surfaces (PESs) and determine the corresponding ground$-$state deformations. For nuclei with odd number of nucleons, a simple blocking method without breaking the time$-$reversal symmetry is adopted \cite{Ring1996}.\par

In these non$-$linear version of our calculations, we use a delta force, i.e., $V = -V_{0} \delta(r)$ with the strength $V_{0} = 350$ $ MeV fm^3$ for the pairing interaction, which has been used in Refs. \cite{Yadav2004,saxenaPLB2019} for the successful description of drip$-$line nuclei. Apart from its simplicity, the applicability and justification of using such a $\delta-$function form of interaction has been discussed in Ref. \cite{Dobaczewski1983}, whereby it has been shown in the context of HFB calculations that the use of a delta force in a finite space simulates the effect of finite range interaction in a phenomenological manner (see also \cite{Bertsch1991} for more details). The computer code used for these calculations is mainly based on the computer code given by Ring \textit{et al.} \cite{ringcpc1997}. The parameter set used is NL3* \cite{Lalazissis09}, which has been widely used to provide an excellent description of spherical as well as in deformed nuclei. For further details of these formulations we refer the reader to Refs. \cite{Boguta77,Boguta83,Furnstahl97,Lalazissis09,Gambhir1989,Singh2013,Geng2003}.\par

For a comparison, in analogy with meson$-$exchange model (DD$-$ME) \cite{Lalazissis05}, we have used density$-$dependent point coupling interaction (DD$-$PC) \cite{niksic08} in the RMF calculations. This variant accurately describes the nuclear ground state properties including the neutron$-$skin thickness, as well as the isoscalar giant monopole resonance excitation energies and dipole polarizabilities. The effective lagrangian for this model in terms of nucleonic field can be expressed as:
\begin{eqnarray} \label{eq:Lagrangian3} 
{\cal L}=&& \overline{\psi}(\imath\gamma . \partial - M)\psi\nonumber\\
&&-\frac{1}{2}\alpha_{S}(\rho)(\overline{\psi}\psi)(\overline{\psi}\psi) -
\frac{1}{2}\alpha_{V}(\rho)(\overline{\psi}\gamma^{\mu}\psi)(\overline{\psi}\gamma_{\mu}\psi)\nonumber
\\ &&- \frac{1}{2}\alpha_{TV}(\rho)(\overline{\psi}\overrightarrow{\tau}\gamma^{\mu}\psi)(\overline{\psi}\overrightarrow{\tau}\gamma_{\mu}\psi) - \frac{1}{2}\delta_{S}(\partial_{\nu}\overline{\psi}\psi)(\partial^{\nu}\overline{\psi}\psi)\nonumber\\
&& - \epsilon\overline{\psi}\gamma . \bf{A}\frac{(1-\tau_3)}{2}\psi 
\end{eqnarray}

Here, the free nucleonic term contains isoscalar$-$scalar (S), isoscalar$-$vector (V) and isovector$-$vector (TV) interactions. The coupling constants $\alpha_{i}(\rho)$ are density dependent and employed as:
\begin{equation} \label{eq:Equation4} 
\alpha_{i}(\rho)  =  a_{i} + (b_{i} + c_{i}x)e^{-d_{i}x},    for \,\,\, i = S, V, TV
\end{equation}
where $x = \rho/\rho_{o}$, and $\rho_{o}$ denotes the nucleon density in symmetric nuclear matter at saturation point. For this density dependent model, we use the TMR separable pairing force of Ref. \cite{TMR} for the short range correlations. This kind of separable pairing force has been adjusted to reproduce the pairing gap of the Gogny force D1S in symmetric nuclear matter. Both forces are of finite range and therefore they show no ultraviolet divergence and do not depend on a pairing cut$-$off. They provide a very reasonable description of pairing correlations all over the periodic table with a fixed set of parameters. In the $^{1}$S$_{0}$ channel the gap equation is given by
 \begin{equation}
\Delta(k) = \int_{0}^{\infty} \frac{k'^2 dk'}{2\pi^2} \langle k|V^{^{1}S_{0}}|k'\rangle \frac{\Delta k'}{2E(k')}
  \end{equation}
  and the pairing force separable in momentum space is
 \begin{equation}
  \langle k|V^{^{1}S_{0}}|k'\rangle = - Gp(k)p(k')
  \end{equation}
The two parameters determining the force are, the pairing strength G and $\alpha$ that goes in the Gaussian ansatz $p(k)= e^{-\alpha^2 k^2}$. Their value has been adjusted to G = 728 MeV fm$^3$ and $\alpha$ = 0.644 fm in order to reproduce the density dependence of the gap at the Fermi surface, calculated with the D1S parametrization of the Gogny force \cite{berger1991}. For these calculations, we have used computer code from Ref. \cite{niksiccpc2014} and the parameter DD$-$PC1 from Ref. \cite{niksic08}.\par

\section{Results and Discussion}
\label{sec:Results and Discussion}
The experimental observables most closely related to the nuclear shape are the quadrupole moments of the excited states and electromagnetic transition rates between them and their measurements that are used for the study ~\cite{Wood,Julin} of shapes and shape coexistence which, in turn, also provide a testing ground for the theoretical nuclear structure model predictions. To explore shape phase transitions and coexisting states in Mo and Ru isotopes, we first locate energy minima with respect to Nilsson deformation parameters ($\beta_{2}$, $\gamma$) and present a complete trace of equillibrium deformations and shapes using NSM and RMF(NL3*) ranging from the first unbound nucleus (even $^{78}$Mo, odd $^{79}$Mo), (even $^{84}$Ru, odd $^{83}$Ru) along the proton drip line to the first unbound nucleus (odd $^{125}$Mo, even $^{144}$Mo), (odd $^{127}$Ru, even $^{152}$Ru) on the neutron drip line. It is to be noted here that first unbound even$-$even nucleus takes it much longer to reach neutron drip line as compared to proton drip line where even and odd first unbound nuclei are close to each other. In case of neutron dripline, the even A nuclei remain bound upto a much higher neutron number as compared to odd A nucleus which is unbound at a much lower neutron number, indicating  strong binding with pairing in e$-$e nuclei. The entire isotopic range of Mo and Ru is well deformed with the predominant oblate and triaxial shapes. \par

\begin{figure}[h]
\centering
\includegraphics[width=0.75\textwidth]{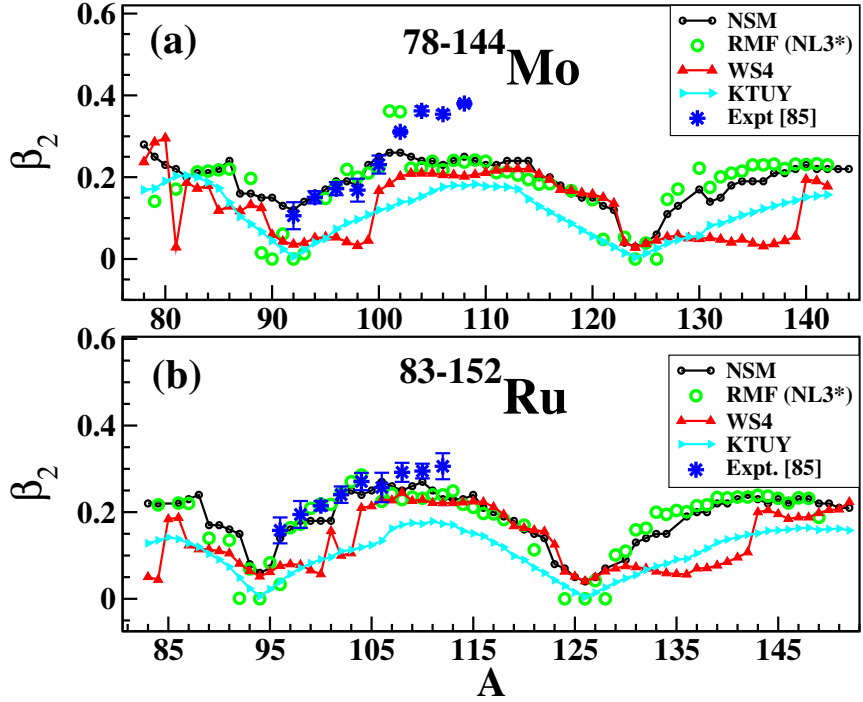}
\caption{Deformation Parameter $\beta_{2}$ vs. A for (a) Mo and (b) Ru isotopes ranging from proton drip line to neutron drip line.}
\label{gdefmoru}
\end{figure}
\hfill
\begin{figure}[h]
\centering
\includegraphics[width=0.75\textwidth]{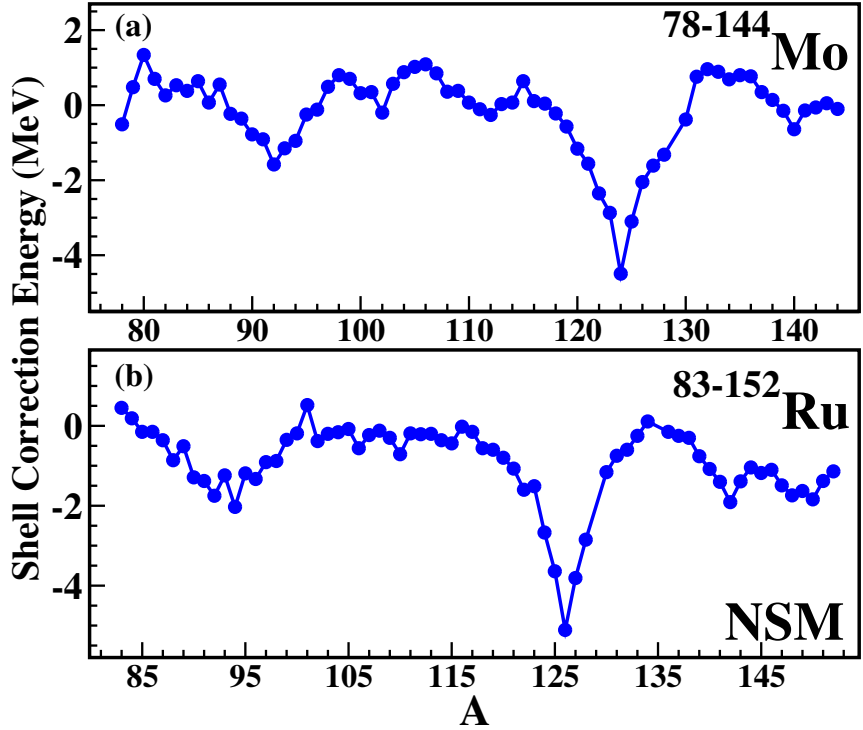}
\caption{Shell correction to energy $\delta$E$_{shell}$ for (a) Mo and (b) Ru isotopes evaluated using NSM.}
\label{ShellCorrc}
\end{figure}

Fig. \ref{gdefmoru} (a) and (b) present the variation of equilibrium deformation vs. A for Mo and Ru isotopic chains. Our computed values using NSM and RMF(NL3*) match well with the available data ~\cite{Wapstra} and seems to be better than the other theoretical predictions of different models viz. Weizsäcker$-$Skyrme (WS4) \cite{ws4} and KTUY \cite{ktuy2005}. \par

The sensitive interplay of the macroscopic bulk properties and microscopic shell structure has been addressed by the inclusion of shell correction and deformation effects ~\cite{MAPLB} using the NSM. The  accurate estimate of the shell correction energy ~\cite{Manpald} is crucial for the precise determination of binding energy, level densities and the other structural properties of the nuclear systems. Computed values of shell correction to energy of the ground$-$state for Mo and Ru isotopic chain is presented in Fig. \ref{ShellCorrc}. Shell correction to energy $\delta$E$_{shell}$ for both Mo and Ru nuclei varies from few KeV to a minima of about $-5$ MeV and $-2$ MeV at N $=$ 82 and 50 respectively. The magic character at N $=$ 50 seems to be slightly weakened due to significant deformation in $^{92}$Mo which shows good agreement with data ~\cite{Raman,Erhard} unlike other theoretical predictions ~\cite{Dhiman,Abusara,Nabi} that predict very small or zero deformation, and thus proves the efficacy of our theoretical model. But since it is not a drip line nucleus, the weakening of magicity is not expected and this calls for further investigation which showed that this nucleus exhibits shape coexistence with a second minima at almost zero deformation. This points towards the shape instability and shape mixing which is seen even at a magic number. \par

\begin{figure}[htb]
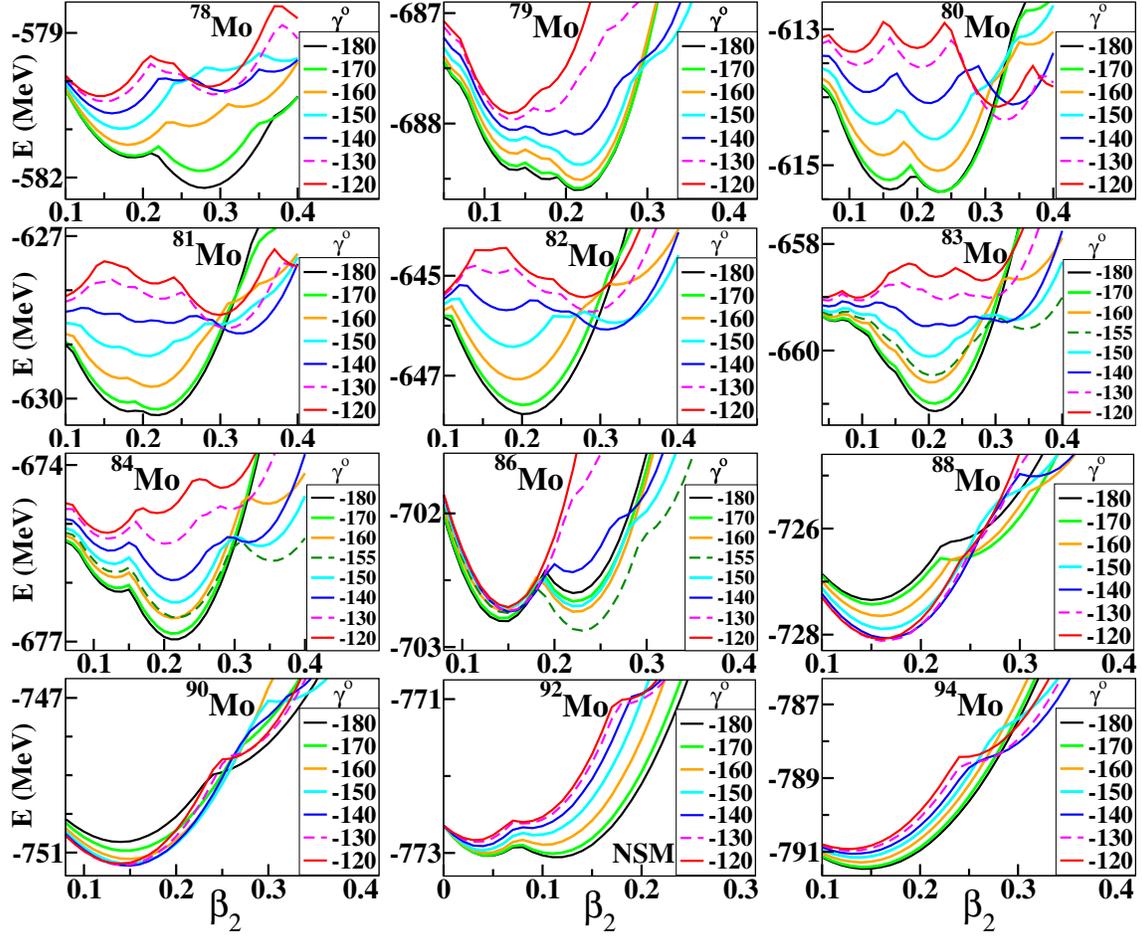

\centering
\begin{subfigure}[t]{0.325\linewidth}
\centering
\includegraphics[width=\linewidth]{gd78mo.eps}
\end{subfigure}
\begin{subfigure}[b]{0.325\textwidth}
\centering
\includegraphics[width=\textwidth]{gd79mo.eps}
\end{subfigure}
\begin{subfigure}[b]{0.325\textwidth}
\centering
\includegraphics[width=\textwidth]{gd80mo.eps}
\end{subfigure}
\begin{subfigure}[b]{0.325\textwidth}
\centering
\includegraphics[width=\textwidth]{gd81mo.eps}
\end{subfigure}
\begin{subfigure}[b]{0.325\textwidth}
\centering
\includegraphics[width=\textwidth]{gd82mo.eps}
\end{subfigure}
\begin{subfigure}[b]{0.325\textwidth}
\centering
\includegraphics[width=\textwidth]{gd83mo.eps}
\end{subfigure}
\begin{subfigure}[b]{0.325\textwidth}
\centering
\includegraphics[width=\textwidth]{gd84mo.eps}
\end{subfigure}
\begin{subfigure}[b]{0.325\textwidth}
\centering
\includegraphics[width=\textwidth]{gd86mo.eps}
\end{subfigure}
\begin{subfigure}[b]{0.325\textwidth}
\centering
\includegraphics[width=\textwidth]{gd88mo.eps}
\end{subfigure}
\begin{subfigure}[b]{0.325\textwidth}
\centering
\includegraphics[width=\textwidth]{gd90mo.eps}
\end{subfigure}
\begin{subfigure}[b]{0.325\textwidth}
\centering
\includegraphics[width=\textwidth]{gd92mo.eps}
\end{subfigure}
\begin{subfigure}[b]{0.325\textwidth}
\centering
\includegraphics[width=\textwidth]{gd94mo.eps}
\end{subfigure}
\caption{Tracing energy minima to locate ground state shape and deformation for  $^{78-94}$Mo isotopes lying close to proton drip line using NSM.}
\label{78to94Mo}
\end{figure}

Fig. \ref{78to94Mo} shows the energy minima traced for the ground states of proton rich Mo isotopes as a function of $\beta_{2}$ and $\gamma$ for $^{78-94}$Mo using NSM. Here we see a variety of shape changes from a well$-$defined oblate ($\gamma$ $=$ $-$180$^o$) minima at the first unbound even$-$even nucleus $^{78}$Mo, to many triaxial $\gamma$ competing with oblate in $^{79}$Mo, to a state where a nearly prolate triaxially soft $\gamma$ $=$ $-$130$^o$ with higher deformation is seen competing with oblate for minima in $^{80}$Mo with an energy difference of around 1 MeV, where, interestingly, one also sees two oblate minima at the same energy but different deformations. With increasing N, the triaxial minima is seen competing more and more strongly resulting in two strong well deformed minima at oblate and triaxial states coexisting at $^{86}$Mo with deeper and strongly deformed triaxial minima. During these rapid shape phase transitions from oblate to triaxial from $^{78}$Mo to $^{86}$Mo, the $\gamma$ states exhibit shape instability and mixing where many $\gamma$s are competing or existing at close energies with different deformations. Further increasing N while moving towards the magic number N $=$ 50, mixing of various shapes mostly triaxial $\gamma$s are seen with no well defined minima. $^{90}$Mo shows lowest minima around $\beta_{2}$ $=$ 0.15 with triaxial shape ($\gamma$ $=$ $-$140$^o$) along with a prolate minima ($\beta_{2}$ $=$ 0.14,) with an energy difference of $\Delta$E = 0.016 MeV, and oblate minima ($\beta_{2}$ = 0.14) with $\Delta$E = 0.614 MeV. This looks like a shape mixing state where all shapes coexist at similar deformation and energy. The closed shell nucleus $^{92}$Mo shows a strong clear minima at oblate shape ($\beta_{2}$ $=$ 0.11, $\gamma$ $=$ $- $180$^o$ ), in good match with experimental data ~\cite{Raman}, along with a second minima at prolate ($\beta_{2}$ $=$ 0.03, $\gamma$ $= -$120$^o$) with energy difference of merely 222 KeV. Although  $^{92}$Mo being a closed shell nucleus is expected to be spherical, but two coexisting states at similar energies make this nucleus interesting for the probes of structural properties. Strong oblate minima is observed at $^{94}$Mo just after the closed shell nucleus $^{92}$Mo. \par

\begin{figure}[htb]
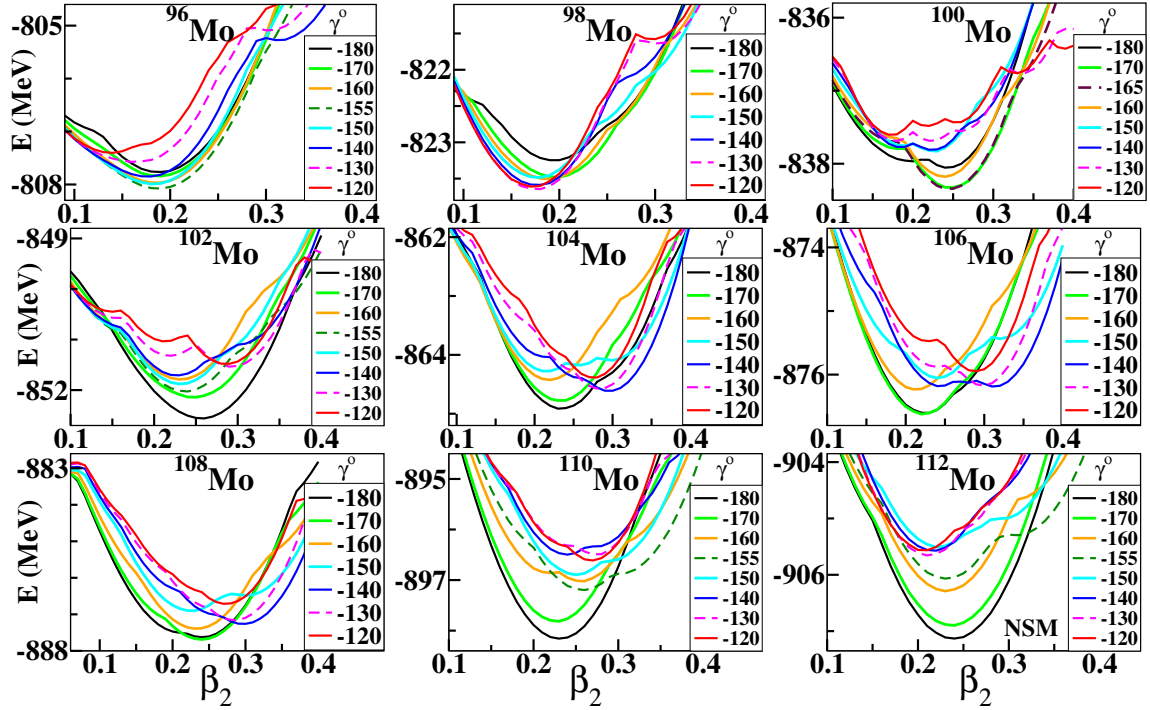

\centering
\begin{subfigure}[b]{0.325\textwidth}
\centering
\includegraphics[width=\textwidth]{gd96mo.eps}
\end{subfigure}
\hfill
\begin{subfigure}[b]{0.325\textwidth}
\centering
\includegraphics[width=\textwidth]{gd98mo.eps}
\end{subfigure}
\begin{subfigure}[b]{0.325\textwidth}
\centering
\includegraphics[width=\textwidth]{gd100mo.eps}
\end{subfigure}
\begin{subfigure}[b]{0.325\textwidth}
\centering
\includegraphics[width=\textwidth]{gd102mo.eps}
\end{subfigure}
\begin{subfigure}[b]{0.325\textwidth}
\centering
\includegraphics[width=\textwidth]{gd104mo.eps}
\end{subfigure}
\begin{subfigure}[b]{0.325\textwidth}
\centering
\includegraphics[width=\textwidth]{gd106mo.eps}
\end{subfigure}
\begin{subfigure}[b]{0.325\textwidth}
\centering
\includegraphics[width=\textwidth]{gd108mo.eps}
\end{subfigure}
\begin{subfigure}[b]{0.325\textwidth}
\centering
\includegraphics[width=\textwidth]{gd110mo.eps}
\end{subfigure}
\begin{subfigure}[b]{0.325\textwidth}
\centering
\includegraphics[width=\textwidth]{gd112mo.eps}
\end{subfigure}
\caption{Tracing energy minima as a function of $\beta_{2}$ and $\gamma$ for stable nuclei $^{96-112}$Mo using NSM.}
\label{96to112Mo}
\end{figure}

Fig. \ref{96to112Mo} presents the curves tracing energy minima as a function of $\beta_{2}$ and $\gamma$ for stable nuclei $^{96}$Mo to $^{112}$Mo. The shape transitions from single well$-$defined and strongly deformed triaxial $^{96}$Mo (with oblate and prolate also lying at very close energies less than 1 MeV and similar deformation) to nearly prolate triaxial $^{98}$Mo to triaxial $^{100}$Mo ($\beta_{2}$ $=$ 0.24) with oblate and prolate shapes lying at close energies with energy difference of a few KeVs. Perhaps this is the reason why this Mo region is known to exhibit rapid shape transition with shape mixing and softness towards triaxiallity. At $^{100}$Mo, the mid shell nucleus with N $=$ 60, one can see a single strong oblate minima with high deformation. Moving further to neutron rich nuclei brings us to the shape coexistence region $^{104-108}$Mo where oblate and triaxial shape coexist with strong deformation in agreement with predictions by Refs. ~\cite{Abusara, Rodriguez} for these nuclei. Interestingly, our calculations show that in $^{104-108}$Mo isotopes, prolate shape also exists along with oblate and triaxial shapes at an energy difference of less than 1 MeV, which matches the predictions of shape coexistence with oblate and prolate shape by the RMF Model where only the oblate and prolate shapes have been considered for the calculations. Our calculations using the NSM theory including triaxiallity, show a unique shape coexistence between oblate, prolate and triaxial shapes which is not seen commonly and hence it would be interesting to probe more into it for better clarity experimentally as well as theoretically. The nuclei $^{110,112}$Mo exhibit shape coexistence with well deformed deeper oblate minima and a shallow triaxial presence at higher deformation  as seen in Fig 4.  \par

\begin{figure}[htbp]
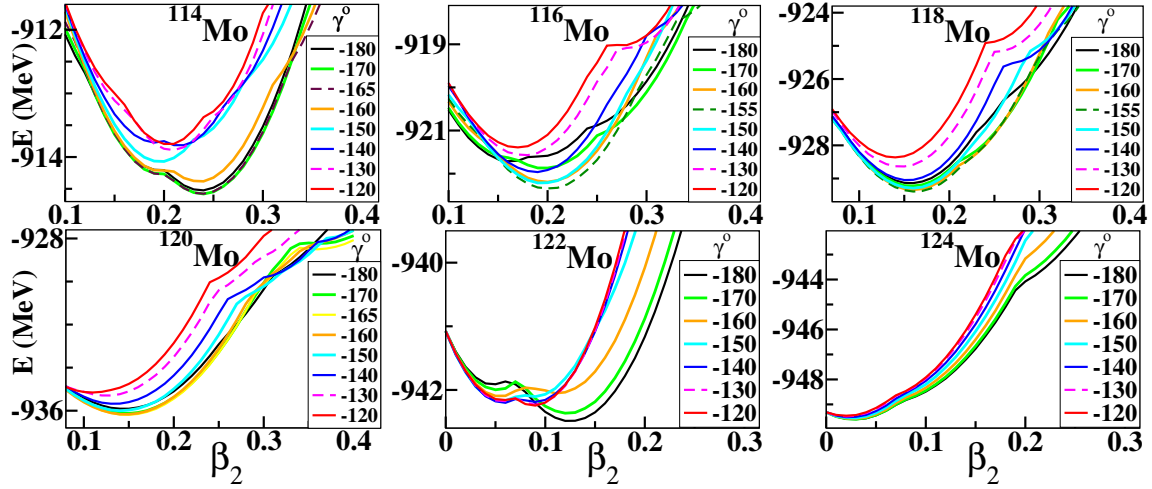

\begin{subfigure}[b]{0.325\textwidth}
\centering
\includegraphics[width=\textwidth]{gd114mo.eps}
\end{subfigure}
\begin{subfigure}[b]{0.325\textwidth}
\centering
\includegraphics[width=\textwidth]{gd116mo.eps}
\end{subfigure}
\begin{subfigure}[b]{0.325\textwidth}
\centering
\includegraphics[width=\textwidth]{gd118mo.eps}
\end{subfigure}
\begin{subfigure}[b]{0.327\textwidth}
\centering
\includegraphics[width=\textwidth]{gd120mo.eps}
\end{subfigure}
\begin{subfigure}[b]{0.327\textwidth}
\centering
\includegraphics[width=\textwidth]{gd122mo.eps}
\end{subfigure}
\begin{subfigure}[b]{0.327\textwidth}
\centering
\includegraphics[width=\textwidth]{gd124mo.eps}
\end{subfigure}
\caption{E vs. $\beta_{2}$ and $\gamma$ for very neutron rich $^{114-124}$Mo isotopes.}
\label{114to124Mo}
\end{figure}

Fig. \ref{114to124Mo} displays the plots of E vs $\beta_{2}$ and $\gamma$ for very neutron rich  nuclei from $^{114}$Mo to $^{124}$Mo which is the first unbound odd nucleus lying at the neutron drip line. This region, close to the neutron drip line, predominantly shows triaxiallity with decreasing $\beta_{2}$ as it is approaching shell closure N$=$82. $^{122}$Mo shows oblate and prolate shape coexistence before it goes to almost spherical shape at closed shell nucleus $^{124}$Mo. \par

\begin{figure}[htb]
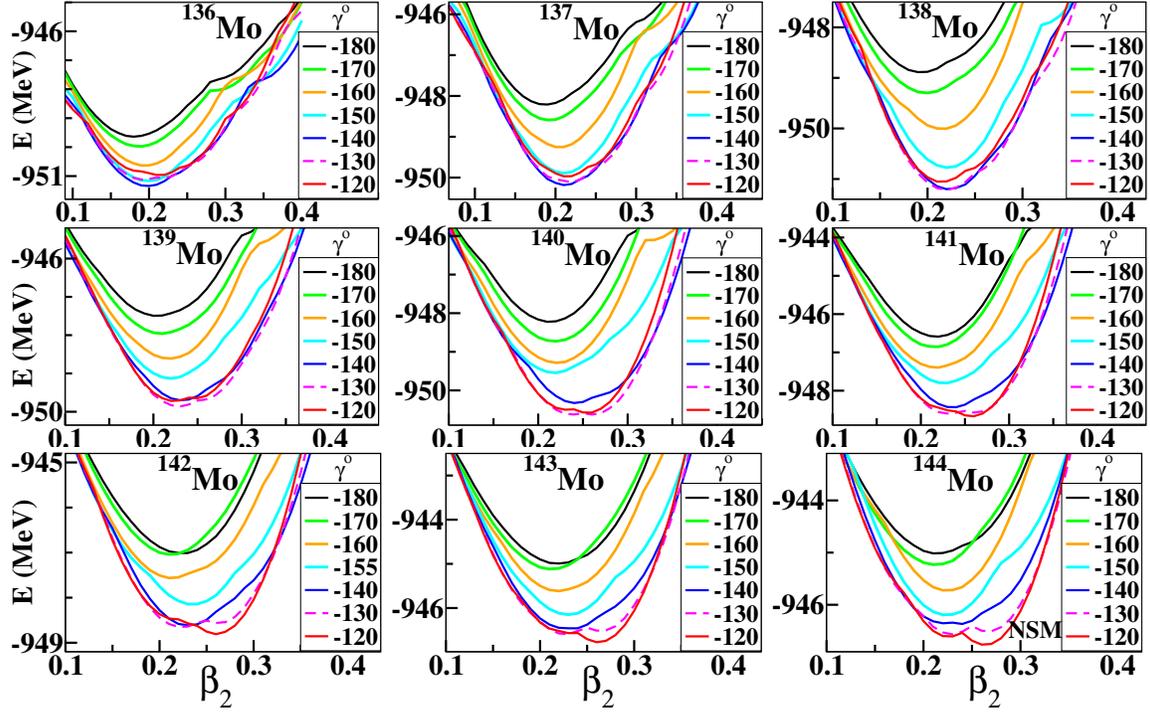

\begin{subfigure}[t]{0.325\textwidth}
\centering
\includegraphics[width=\textwidth]{gd136mo.eps}
\end{subfigure}
\begin{subfigure}[t]{0.325\textwidth}
\centering
\includegraphics[width=\textwidth]{gd137mo.eps}
\end{subfigure}
\begin{subfigure}[t]{0.325\textwidth}
\centering
\includegraphics[width=\textwidth]{gd138mo.eps}
\end{subfigure}
\begin{subfigure}[t]{0.325\textwidth}
\centering
\includegraphics[width=\textwidth]{gd139mo.eps}
\end{subfigure}
\begin{subfigure}[t]{0.325\textwidth}
\centering
\includegraphics[width=\textwidth]{gd140mo.eps}
\end{subfigure}
\begin{subfigure}[t]{0.325\textwidth}
\includegraphics[width=\textwidth]{gd141mo.eps}
\end{subfigure}
\begin{subfigure}[t]{0.327\textwidth}
\includegraphics[width=\textwidth]{gd142mo.eps}
\end{subfigure}
\begin{subfigure}[t]{0.327\textwidth}
\centering
\includegraphics[width=\textwidth]{gd143mo.eps}
\end{subfigure}
\begin{subfigure}[t]{0.327\textwidth}
\centering
\includegraphics[width=\textwidth]{gd144mo.eps}
\end{subfigure}
\caption{E vs. $\beta_{2}$ and $\gamma$ for extremely neutron rich $^{136-144}$Mo isotopes.}
\label{136to144Mo}
\end{figure}

Moving further towards the extremely neutron rich nuclei  $^{136-144}$Mo upto the neutron drip line where an even$-$even nucleus is unbound at $^{144}$Mo in Fig. \ref{136to144Mo}, we find prolate shape as the most dominant shape phase which is found only in this region of the entire chain of Mo isotopes. Shape coexistence is not seen in this region close to neutron drip line. \par

\begin{figure}[htbp]
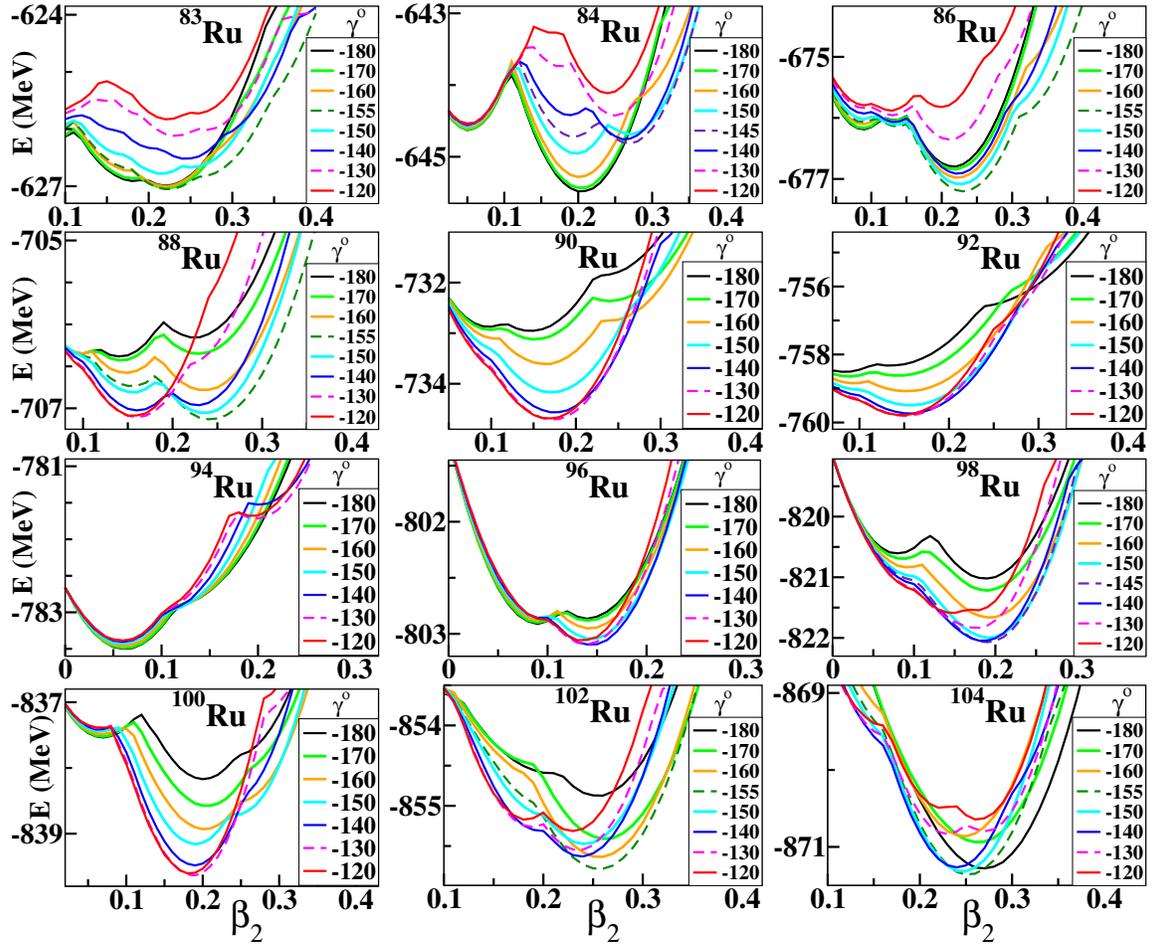

\begin{subfigure}[b]{0.325\textwidth}
\centering
\includegraphics[width=\textwidth]{gd83ru.eps}
\end{subfigure}
\begin{subfigure}[b]{0.325\textwidth}
\centering
\includegraphics[width=\textwidth]{gd84ru.eps}
\end{subfigure}
\begin{subfigure}[b]{0.325\textwidth}
\centering
\includegraphics[width=\textwidth]{gd86ru.eps}
\end{subfigure}
\begin{subfigure}[b]{0.325\textwidth}
\centering
\includegraphics[width=\textwidth]{gd88ru.eps}
\end{subfigure}
\begin{subfigure}[b]{0.325\textwidth}
\centering
\includegraphics[width=\textwidth]{gd90ru.eps}
\end{subfigure}
\begin{subfigure}[b]{0.325\textwidth}
\centering
\includegraphics[width=\textwidth]{gd92ru.eps}
\end{subfigure}
\begin{subfigure}[b]{0.325\textwidth}
\centering
\includegraphics[width=\textwidth]{gd94ru.eps}
\end{subfigure}
\begin{subfigure}[b]{0.325\textwidth}
\centering
\includegraphics[width=\textwidth]{gd96ru.eps}
\end{subfigure}
\begin{subfigure}[b]{0.325\textwidth}
\centering
\includegraphics[width=\textwidth]{gd98ru.eps}
\end{subfigure}
\begin{subfigure}[b]{0.325\textwidth}
\centering
\includegraphics[width=\textwidth]{gd100ru.eps}
\end{subfigure}
\begin{subfigure}[b]{0.33\textwidth}
\centering
\includegraphics[width=\textwidth]{gd102ru.eps}
\end{subfigure}
\begin{subfigure}[b]{0.33\textwidth}
\centering
\includegraphics[width=\textwidth]{gd104ru.eps}
\end{subfigure}
\caption{Tracing E minima obtained by NSM to evaluate ground state shape and deformation for proton drip$-$line nuclei $^{83-104}$Ru}
\label{83to104Ru}
\end{figure}

Fig. \ref{83to104Ru} traces the E minima as a function of $\beta_{2}$ and $\gamma$ for $^{83-104}$Ru isotopes. The first proton unbound odd nucleus $^{83}$Ru shiws deep oblate along with triaxial shapes competing very closely almost coinciding. First proton unbound even$-$even nucleus $^{84}$Ru shows a shape coexistence with a deep oblate minima with two little shallow minima of triaxiallity with am small energy difference, which eventually moves to a deeper well$-$defined triaxial minima in $^{86}$Ru, to  another shape coexisting state with triaxial and a nearly prolate minima in  $^{88}$Ru at almost similar energies. Such type of rapid shape changes are peculiar to this region. The equillibrium deformation is reducing as N is approaching the magic number N $=$ 50. At $^{92}$Ru, prolate and triaxial $\gamma$ are competing for minima and it is difficult to decide a single minima. $^{94}$Ru (N $=$ 50) shows a single minima by all $\gamma$s at $\beta_{2}$ = 0.06. At further  increasing N, one can see shallow coexisting triaxial and prolate or nearly prolate shapes in  $^{96-102}$Ru. However $^{104}$Ru shows coexisting oblate and triaxial minima with very close energies along with a prolate minima also at $\Delta E$ $=$ 0.732 MeV. 

\begin{figure}[htb]
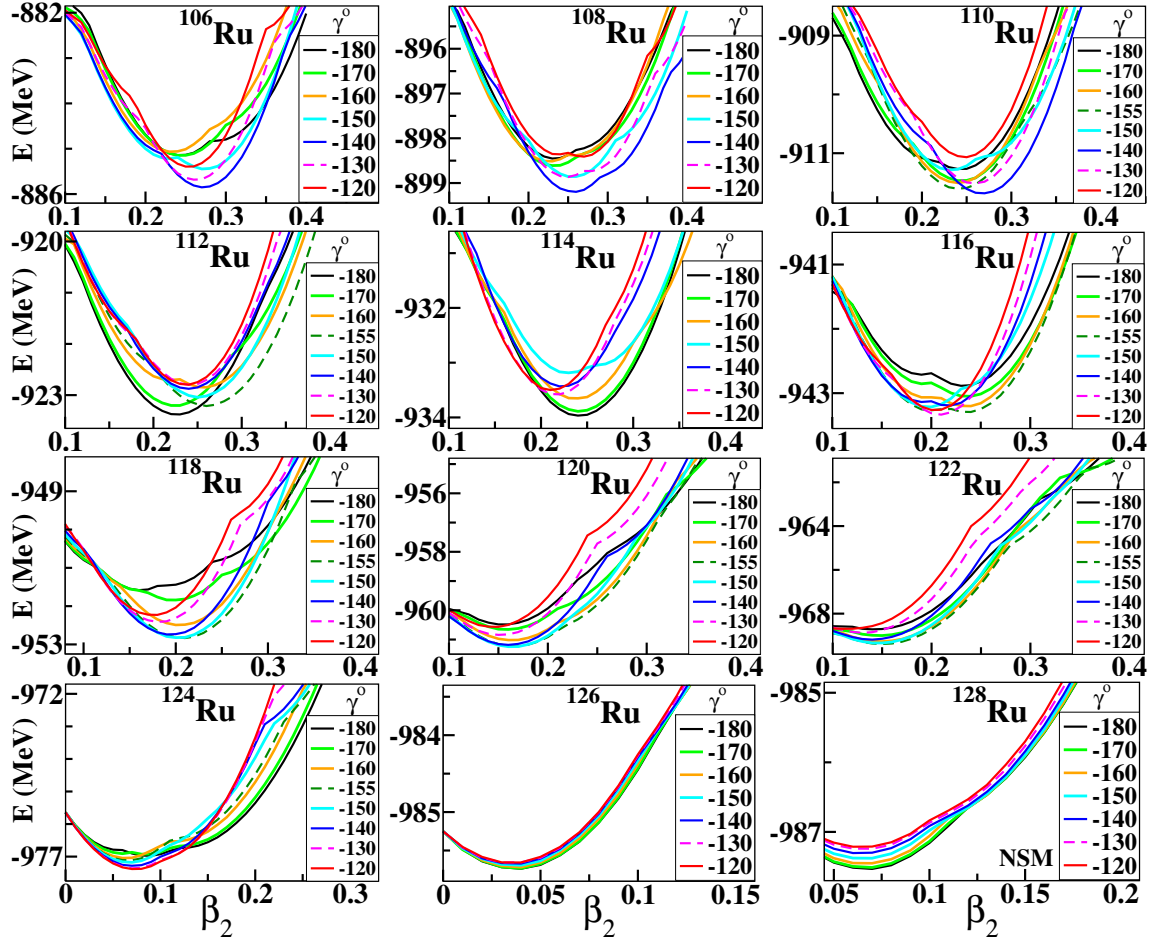

\begin{subfigure}[b]{0.325\textwidth}
\centering
\includegraphics[width=\textwidth]{gd106ru.eps}
\end{subfigure}
\begin{subfigure}[b]{0.325\textwidth}
\centering
\includegraphics[width=\textwidth]{gd108ru.eps}
\end{subfigure}
\begin{subfigure}[b]{0.325\textwidth}
\centering
\includegraphics[width=\textwidth]{gd110ru.eps}
\end{subfigure}
\begin{subfigure}[b]{0.325\textwidth}
\centering
\includegraphics[width=\textwidth]{gd112ru.eps}
\end{subfigure}
\begin{subfigure}[b]{0.325\textwidth}
\centering
\includegraphics[width=\textwidth]{gd114ru.eps}
\end{subfigure}
\begin{subfigure}[b]{0.325\textwidth}
\centering
\includegraphics[width=\textwidth]{gd116ru.eps}
\end{subfigure}
\begin{subfigure}[b]{0.325\textwidth}
\centering
\includegraphics[width=\textwidth]{gd118ru.eps}
\end{subfigure}
\begin{subfigure}[b]{0.325\textwidth}
\centering
\includegraphics[width=\textwidth]{gd120ru.eps}
\end{subfigure}
\begin{subfigure}[b]{0.325\textwidth}
\centering
\includegraphics[width=\textwidth]{gd122ru.eps}
\end{subfigure}
\begin{subfigure}[b]{0.325\textwidth}
\centering
\includegraphics[width=\textwidth]{gd124ru.eps}
\end{subfigure}
\begin{subfigure}[b]{0.327\textwidth}
\centering
\includegraphics[width=\textwidth]{gd126ru.eps}
\end{subfigure}
\begin{subfigure}[b]{0.327\textwidth}
\centering
\includegraphics[width=\textwidth]{gd128ru.eps}
\end{subfigure}
\caption{E vs. $\beta_{2}$ and $\gamma$ for $^{106-128}$Ru isotopes obtained using NSM.}
\label{106to128Ru}
\end{figure}

In Fig. \ref{106to128Ru}, E vs. $\beta_{2}$ and $\gamma$ curves show the ground state shapes and deformations of Ru isotopes from A $=$ 106 to the neutron drip$-$line at A $=$ 128. The $^{106-110}$Ru isotopes show many $\gamma$s competing for a triaxial minima all showing very similar variation. $^{112}$Ru shows shape coexistence between oblate and triaxial with a deeper single oblate minima at $^{114}$Ru. However, prolate shape is also competing for minima here at a small energy difference and slightly different deformation. In $^{116}$Ru, we can see a triaxial minima with prolate competing closely for minima. For $^{118-122}$Ru, triaxial shape predominates. The magic nuclei $^{126}$Ru (N = 82) shows small deformation with all the shapes almost coinciding with oblate showing slightly deeper minima. $^{128}$Ru shows an oblate minima with all other $\gamma$ very closely competing. \par

\begin{figure}[htbp]
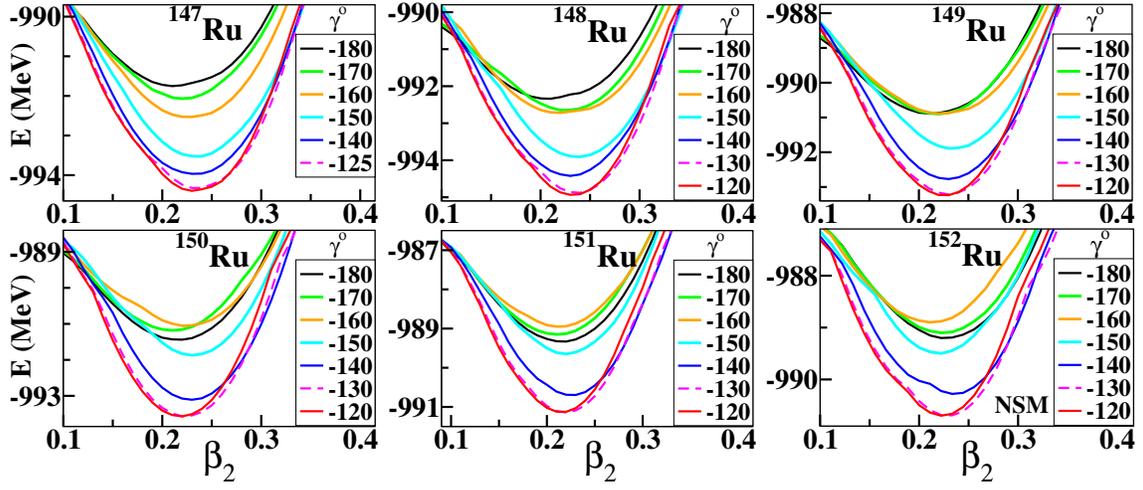

\centering
\begin{subfigure}[b]{0.325\textwidth}
\centering
\includegraphics[width=\textwidth]{gd147ru.eps}
\end{subfigure}
\begin{subfigure}[b]{0.325\textwidth}
\centering
\includegraphics[width=\textwidth]{gd148ru.eps}
\end{subfigure}
\begin{subfigure}[b]{0.325\textwidth}
\centering
\includegraphics[width=\textwidth]{gd149ru.eps}
\end{subfigure}
\begin{subfigure}[b]{0.325\textwidth}
\centering
\includegraphics[width=\textwidth]{gd150ru.eps}
\end{subfigure}
\begin{subfigure}[b]{0.325\textwidth}
\centering
\includegraphics[width=\textwidth]{gd151ru.eps}
\end{subfigure}
\begin{subfigure}[b]{0.325\textwidth}
\centering
\includegraphics[width=\textwidth]{gd152ru.eps}
\end{subfigure}
\caption{E vs. $\beta_{2}$ and $\gamma$ for extremely neutron rich Ru isotopes from $A=147$ to neutron drip$-$line $A=152$ calculated using NSM.}
\label{147to152Ru}
\end{figure}

Fig. \ref{147to152Ru} shows the ground state energy minima for extremely neutron rich Ru isotopes as a function of $\beta_{2}$ and $\gamma$ from $^{147}$Ru upto the neutron drip$-$line nucleus $^{152}$Ru, the first neutron unbound even$-$even nucleus on neutron drip line. A well$-$defined prolate ($\gamma$ $= -$120$^o$) shape phase is seen in this region at around $\beta_{2}$ $\approx$ 0.22. \par

\begin{figure}[htbp]
\centering
\frame{\includegraphics[width=0.8\textwidth]{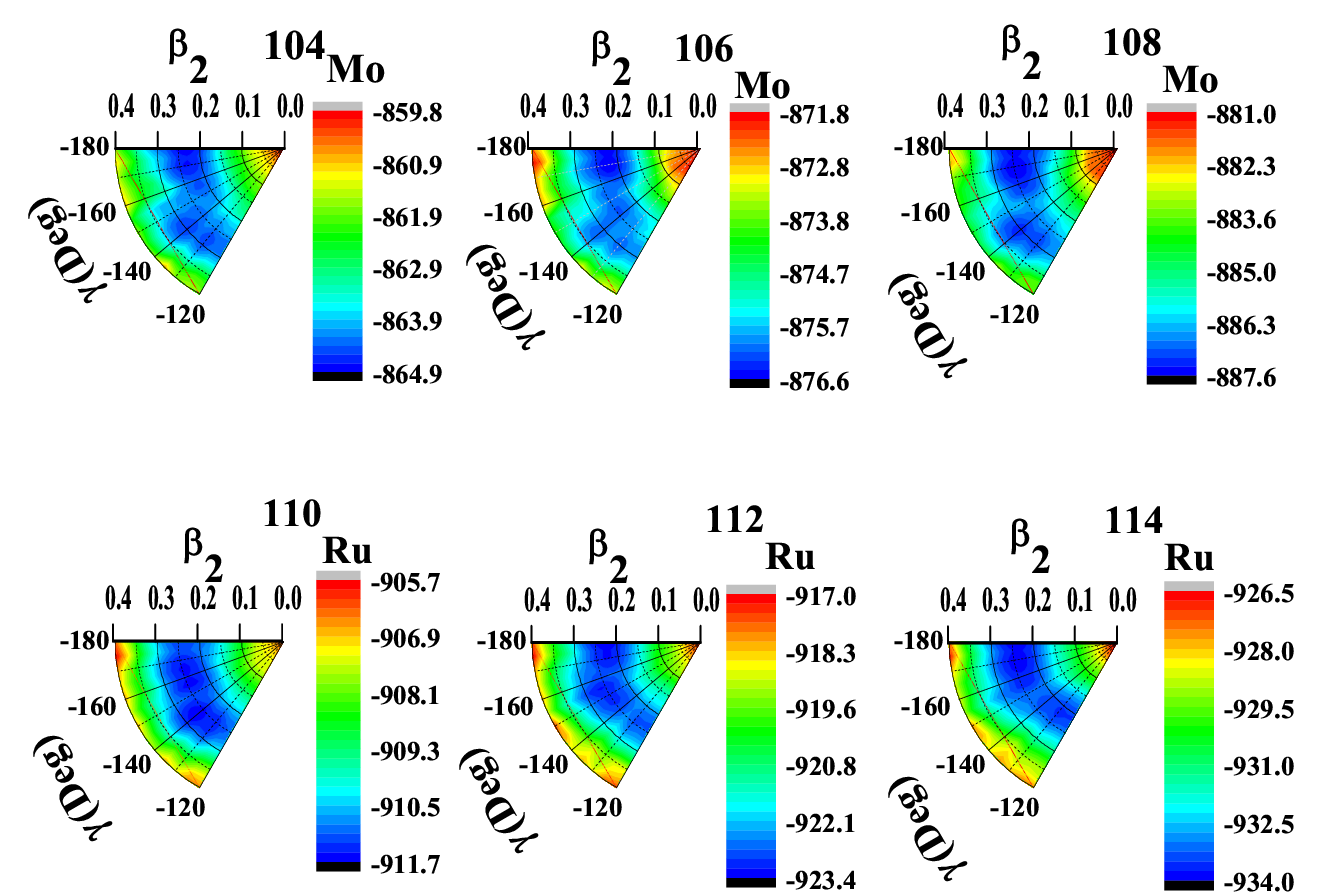}}
\caption{Potential energy surfaces drawn for the shape coexisting nuclei $^{104,106,108}$Mo and $^{110,112,114}$Ru.}
\label{SurfPlot}
\end{figure}

As is evident from Figs. \ref{78to94Mo} to \ref{147to152Ru} that they are presenting a unique and novel way used by us to show the energy minima tracing the shape and deformation of nuclei as a function of  $\beta_{2}$ and $\gamma$ as derived from our NSM calculations. This representation shows the exact value of energy and its variation with each  $\beta_{2}$ and $\gamma$ value and helps in tracing the E minima as well as the shape coexistence in a very easy and decisive manner.  Various $\gamma$ competing for minima is also seen very clearly along with one or two minima. However, since the plots of potential energy surfaces are a popular way to show the shape evaluation, we also plot potential energy surfaces in Fig. \ref{SurfPlot} for a few nuclei that are predicted to exhibit shape coexistence by both the theories NSM and RMF in a more conventional way. \par

\begin{figure}[htbp]
\centering
\includegraphics[width=0.8\textwidth]{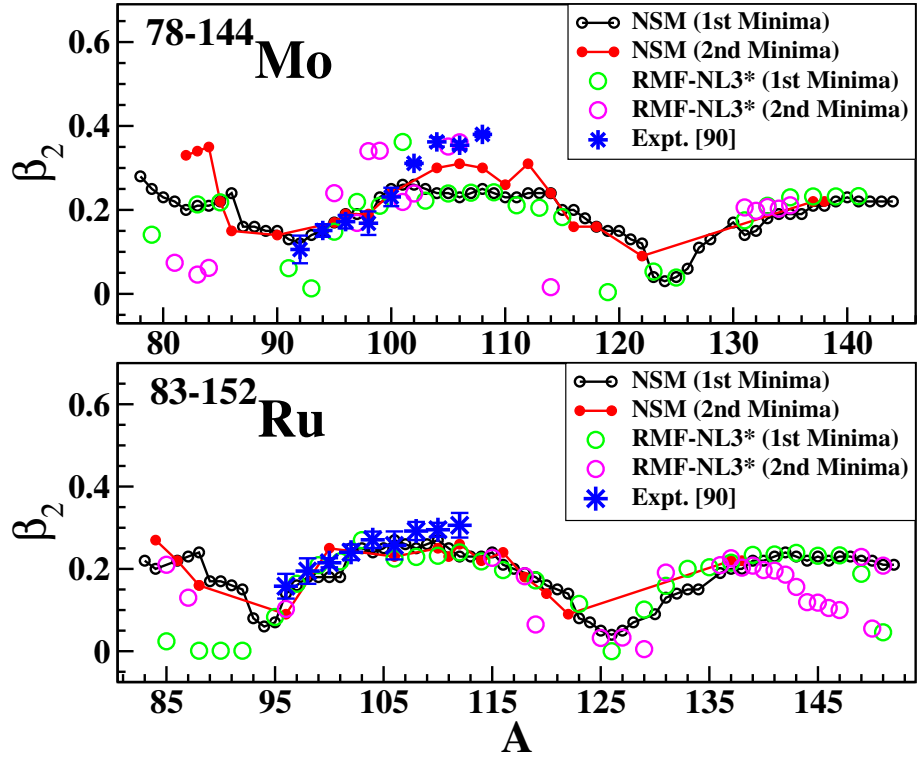}
\caption{Deformation Parameter $\beta_{2}$ for Mo and Ru isotopes corresponding to first and second E minima in shape coexisting nuclei.}
\label{secondminima}
\end{figure}

Fig. \ref{secondminima} compares the values of the equillibrium deformation $\beta_{2}$ obtained by the first minima i.e. the deepest energy minima, and second energy minima in shape coexisting nuclei, along with the available experimental data ~\cite{Raman}. The  $\beta_{2}$ value corresponding to the first energy minima matches well with the data as also shown in Fig. \ref{gdefmoru} except for few nuclei. However, interestingly the $\beta_{2}$ values predicted by the second minima (as seen in Fig. \ref{secondminima}) match better with the experimental values ~\cite{Raman} than that predicted by the first minima in few nuclei ($^{104-108}$Mo) which exhibit shape coexistence. This indicates the possibility of the nucleus being in the second minima state with some finite lifetime during the probe leading to better match with the data. This further invokes the investigation on the lifetimes of the coexisting states, decay modes and the stability of the nucleus as a whole on the theoretical as well as the experimental front.  \par

\begin{figure}[htbp]
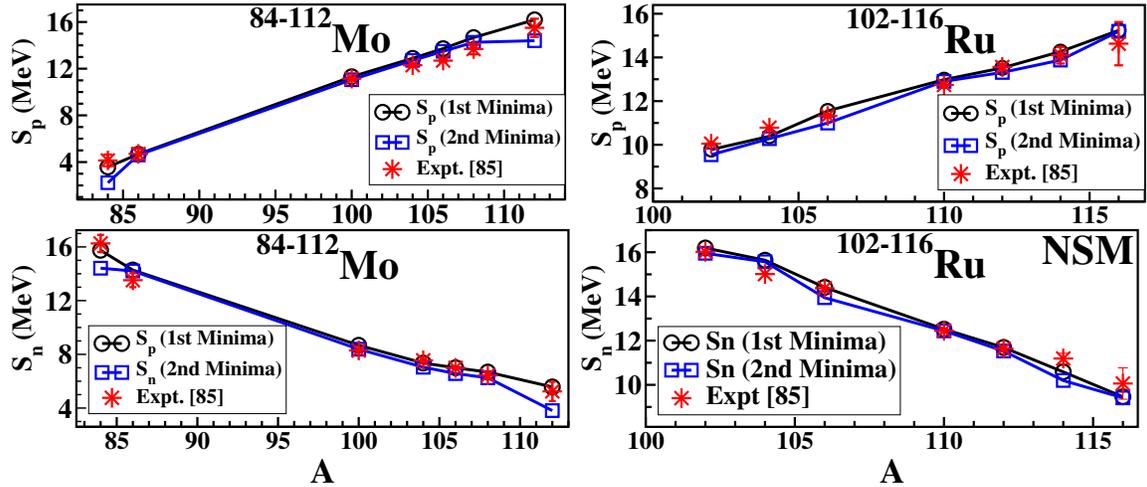

\begin{subfigure}[b]{0.49\textwidth}
\centering
\includegraphics[width=\textwidth]{mosp.eps}
\end{subfigure}
\begin{subfigure}[b]{0.49\textwidth}
\centering
\includegraphics[width=\textwidth]{rusp.eps}
\end{subfigure}
\begin{subfigure}[b]{0.493\textwidth}
\centering
\includegraphics[width=\textwidth]{mosn.eps}
\end{subfigure}
\begin{subfigure}[b]{0.493\textwidth}
\centering
\includegraphics[width=\textwidth]{rusn.eps}
\end{subfigure}
\caption{$S_{p}$ and S$_{n}$ calculated from first and second energy minima for shape coexisting nuclei evaluated using NSM. }
\label{SpSn}
\end{figure}

In view of the above, we plot 1-neutron separation energy (S$_n$) and 1-proton separation energy (S$_p$)  calculated by NSM using the first and second minima of shape coexisting nuclei in Fig. \ref{SpSn}. It is interesting to note that the proton separation energy for a few nuclei ($^{86,104,106,108,112}$Mo) calculated from the energies corresponding to second minima show better agreement with the experimental values~\cite{Wapstra} than that calculated using the first minima. The same is observed for neutron separation energy calculations in $^{86, 100, 106}$Mo nuclei. Incidently, the experimental values ~\cite{Raman} of ground state deformation, that are in better agreement with deformations corresponding to the second minima for $^{100,104,106,108}$Mo nuclei shown in Fig. \ref{secondminima}, also show better agreement of separation energies with the experimental values. This is an unusual observation which is indicating a possibility of nucleus spending some time in second minima state during measurements before decaying or going to the ground state corresponding to the first minima. However, validation of these observations and our speculations needs more investigation although there is no taboo for this possiblity to be true. Our speculation on shape coexistence impacting the lifetimes as also suggested in our earlier work ~\cite{GMAJJPGlife} emphasizes the importance of shape coexistence and the investigation on second minima states that may provide some lifetime sufficient for probes before decay in case of proton and neutron emitters near the drip lines ~\cite{GMAJOGMG,GMAJJPGlife}. Therefore, a better match of deformation and separation energy with those corresponding to second energy minima in shape coexisting nuclei, as observed in this work,  is pointing towards  the influence of shape coexistence on structural properties and lifetimes which may be useful for probes.\par

We  also perform calculations by using the RMF theory to search for the shape coexisting nuclei in Mo and Ru isotopes. Table \ref{rmf-comp} presents our calculated values of the deformation and energy using the first and second minima obtained by RMF calculations using non$-$linear version NL3$^{*}$ parameter \cite{lala,bhuyan2015,Abusara,shi2018} and the density$-$dependent point coupling interaction (DD$-$PC) \cite{niksic08}. The values of deformation using non$-$linear version and density dependent version are given in the Table \ref{rmf-comp} along with the corresponding energy of first and second minima. The difference between the energies of first and second minima ($\Delta$E) is also included in the Table. It is gratifying to note that the phenomenon of shape coexistence in the considered nuclei holds firmly even using different variants of RMF. Both the versions of RMF i.e. NL3* and DD$-$PC result in the same deformation for both the minima (please see the Table  \ref{rmf-comp}). The excellent match in the results of the RMF with two variants, demonstrates the validity of our results and reaffirms the model independency which is a greatly needed criterion to establish outcome of this study related to the half$-$lives. 

\begin{table}[!htbp]
\caption{Deformation and Energy of first and second minima obtained with RMF calculations using various force parameters. The sixth and eleventh columns ($\Delta$E) show difference between the energies between first and second minima. }
\centering
\resizebox{0.95\textwidth}{!}{%
\begin{tabular}{|c|c|c|c|c|c|c|c|c|c|c|}
\hline
\multicolumn{1}{|c|}{Nucleus} &
\multicolumn{5}{c|}{NL3*}&
\multicolumn{5}{c|}{DD-PC1}\\
\hline
\multicolumn{1}{|c|}{} &
\multicolumn{2}{c|}{First Minima} &
\multicolumn{2}{c|}{Second Minima}&
\multicolumn{1}{c|}{$\Delta$E}&
\multicolumn{2}{c|}{First Minima} &
\multicolumn{2}{c|}{Second Minima}&
\multicolumn{1}{c|}{$\Delta$E}\\
\cline{2-5} \cline{7-10}
\multicolumn{1}{|c|}{} &
\multicolumn{1}{c|}{$\beta_{2}$} &
\multicolumn{1}{c|}{E$_1$ (MeV)} &
\multicolumn{1}{c|}{$\beta_{2}$} &
\multicolumn{1}{c|}{E$_2$ (MeV)} &
\multicolumn{1}{c|}{(MeV)} &
\multicolumn{1}{c|}{$\beta_{2}$} &
\multicolumn{1}{c|}{E$_1$ (MeV)} &
\multicolumn{1}{c|}{$\beta_{2}$} &
\multicolumn{1}{c|}{E$_2$ (MeV)} &
\multicolumn{1}{c|}{(MeV)}\\
\hline
$^{104}$Mo  &  -0.2	&-882.84	&0.6&	-882.44	&0.40&	-0.2&	-885.50&	0.6&	-885.31	&0.19 \\
$^{106}$Mo  &  -0.2	&-894.98	&0.4&	-894.05	&0.93&	-0.2&	-898.03&	0.4&	-897.00	&1.03 \\
$^{108}$Mo  &  -0.2	&-906.32	&0.4&	-905.02	&1.30&	-0.2&	-909.65&	0.4&	-908.46	&1.19 \\
$^{110}$Ru  &  -0.2	&-930.16	&0.3&	-928.76	&1.40&	-0.2&	-932.82&	0.3&	-931.55	&1.27 \\
$^{112}$Ru  &  -0.2	&-952.82	&0.3&	-950.43	&2.39&	-0.2&	-945.04&	0.3&	-943.50	&1.54 \\
\hline
\end{tabular}}
\label{rmf-comp}
\end{table}

\begin{figure}[h!]
\centering
\includegraphics[width=0.9\textwidth]{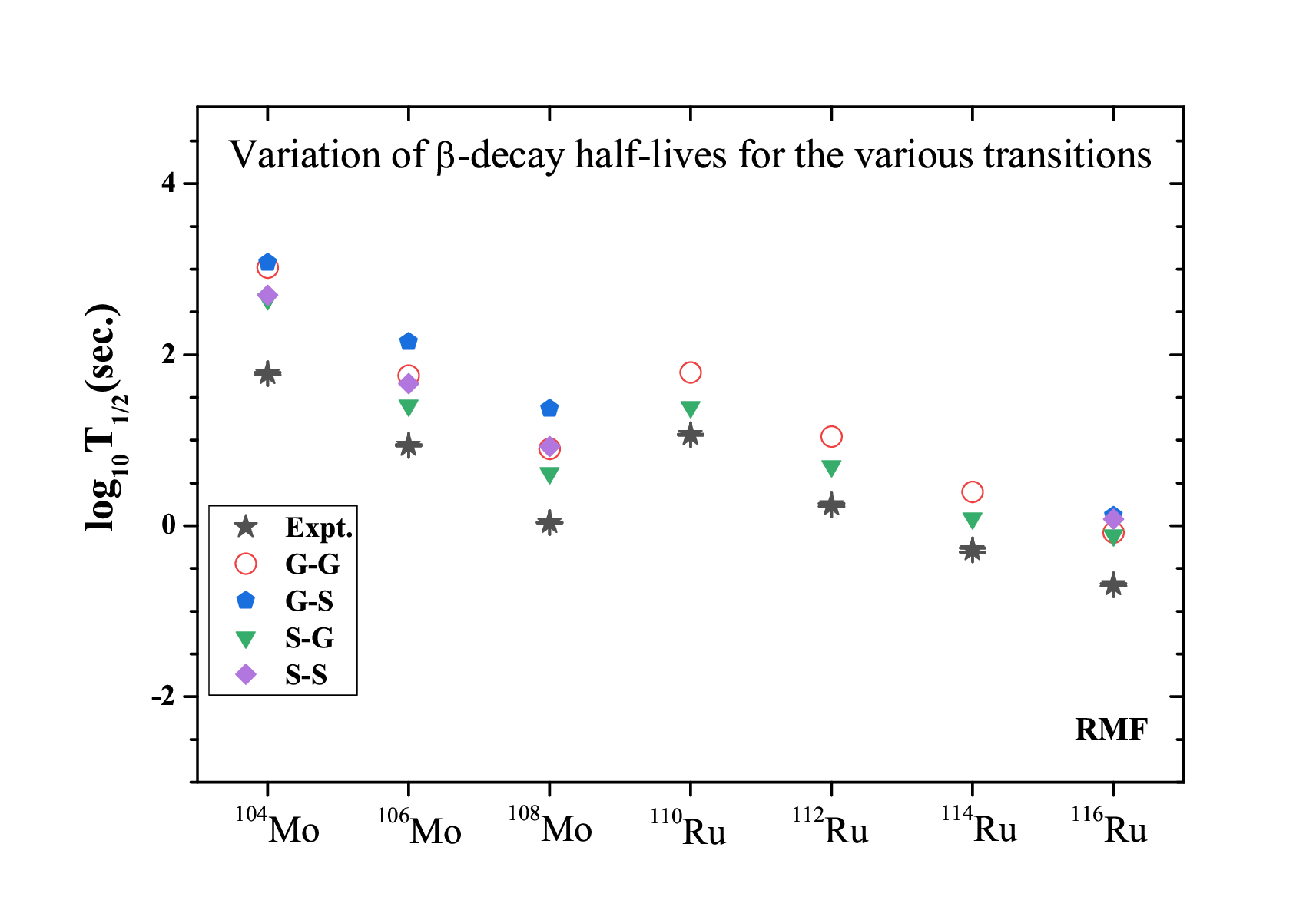}
\caption{$\beta-$decay half$-$lives for selected shape coexisting Mo and Ru nuclei for various transitions as found from RMF(NL3*) calculations.}
\label{halflives}
\end{figure}

The estimation of $\beta-$decay half$-$live, which is  one of the most experimentally accessible physical quantity, is very crucial for the astrophysical processes ~\cite{Yoshida} as it plays a decisive role in determining the time scale of the r$-$process nucleosynthesis~\cite{Nishimura2012}. Observations have shown that the short half$-$lives around A $\approx$ 110 region influence the r$-$process~\cite{Nishimura2011} and the nuclear deformation plays an important role in the $\beta-$decay half$-$lives. Hence, the effect of shape coexistence could be significant in determining the half$-$lives due to the transition from the one shape of the parent to the one shape of the daughter and the combinations if both are shape isomers. For better insights into this, we have examined and computed $\beta-$decay half$-$lives of a few representative isotopes of Mo and Ru which show shape coexistence with both NSM and the RMF theory \cite{saxenaPLB2019,saxena2017,saxenaplb2017}. The corresponding calculations of lifetime  are performed for both the shapes of the daughter nuclei. For simplicity, we have selected a recent empirical formula that is mainly dependent on the Q$-$value and modified by Sobhani et al. \cite{sobhani2022} for the calculations of $\beta-$decay half$-$lives. We have considered 4 allowed transitions which include the first minima (parent) to first minima (daughter) referred to as G$-$G, the second minima of parent to the first minima of the daughter referred to as S$-$G, the first minima of parent to second minima of the daughter (G$-$S), and second minima of parent to second minima of daughter referred as S$-$S. The corresponding half$-$lives are shown in Fig. \ref{halflives} for a few selected isotopes of Mo and Ru as mentioned earlier. The influence of shape coexistence is evident in $\beta-$decay half$-$lives, where G$-$G transition half$-$lives are quite far from the experimental points in few nuclei as compared to the other transitions like  S$-$G, S$-$S. This clearly indicates the effect of shapes on the lifetime estimation. It is to be noted here that it may not be neccessary that G$-$G transitions do not match with the experimental data, it only indicates that S$-$G and S$-$S transitions in the shape coexisting nuclei are possibly influencing the lifetimes sometimes especially when G$-$G transitions are not matching well. This crude estimation is just to show the change in the half$-$lives with the shapes in a qualitative manner although we got quite meaningful and useful results in this work. However, more rigorous calculations using quasi$-$particle random$-$phase approximation (QRPA) or other theoretical work would be one step ahead to get more precise perspectives. Although this work is our preliminary attemp in this direction, but it certainly shows a significant influence of shape coexisting states on the structural properties, stability and lifetimes which is very promising and encouraging to continue to make more efforts to get more information and better insights. \par

\section{Conclusion}
A comprehensive study to trace the shape coexistence in astrophysically interesting  Mo and Ru isotopes has been conducted. This region known for the rapid shape phase transition, shape instabilities and coexistence, has been studied for the entire isotopic chain of Mo and Ru from proton drip line to neutron drip line, within the microscopic theoretical framework using Nilsson Strutinsky Method and Relativistic Mean Field Theory. Isotopic chains of both the nuclei Mo and Ru exhibit rapid shape phase transitions, shape instabilities and mixing and the $\gamma$ softness with the predominance of triaxial shape phase. Since a nucleus with coexisting eigenstates, also exists in a second minima state lying close to the ground state that may have lifetime sufficient for probes, we have explored the structural properties and lifetimes with the first and second minima both. Our calculation show that in few of the shape coexisting nuclei, the structural properties seem to be influenced by the shape coexistence as deformation and separation energy corresponding to the second minima of coexisting state match better than that corresponding to first minima with the experimental data. Further, the $\beta-$decay half$-$lives estimated for a few nuclei exhibiting shape coexistence show that the transitions from second minima are found to be matching better with the data in few of the cases. This clearly indicates the influence of shape coexistence on structural properties, stability and half$-$lives which is a significant outcome of this study. However, it certainly needs more rigorous treatment to get better clarity on the subject.\par

\section{Acknowledgement}
Authors M. Aggarwal and G. Saxena acknowledge the support provided by DST, Govt. of India. and SERB (DST), Govt. of India under SR/WOS$-$A/PM$-$30/2019 and SIR/2022/000566 respectively.
\pagebreak


\begin{thebibliography}{00}
\bibitem{Bethe} H. A. Bethe, Phys. Rev., {\bf{55}}, 434 (1939).
\bibitem{Burbidge} E.M.Burbidge, G. R. Burbidge, W. A. Fowler, F. Hoyle, Rev. Mod. Phys. {\bf{29}}, 547 (1957).
\bibitem{Baade} W. Baade, F. Zwicky, Proc. Natl. Acad. Sci. {\bf{20}}, 254 (1934).
\bibitem{Nabi} J. Nabi, T. Bayram, Astrophys. Space Sci. {\bf {365:19}}, 1  (2020).
\bibitem{Cowan} J. J. Cowan, F.K. Thielemann, and J. W. Truran, Phys. Rep. {\bf{208}}, 267 (1991).
\bibitem{Sasaki} H. Sasaki et al., The Astrophysical Journal, {\bf{924:29}}, 29 (2022).
\bibitem{MAPRC89} M. Aggarwal, Phys. Rev. {\bf{C 89}}, 024325 (2014).
\bibitem{Strut1967} V. M. Strutinsky, Nucl. Phys. {\bf{A 95}}, 420, (1967). 
\bibitem{Strut1968} V. M. Strutinsky, Nucl. Phys. {\bf {A 122}}, 1 (1968).
\bibitem{Nilsson} Sven Gösta Nilsson et al., Nucl. Phys. {\bf {A 131}}, 1 (1969)
\bibitem{MollerNix} P. Möller  and J. R. Nix, J. of Phys. G: Nucl. Part. Phys. {\bf {20}}, 1681 (1994).
\bibitem{Yoshida} K. Yoshida, Y. Niu, and F. Minato, Phys. Rev. {\bf{C 108}}, 034305 (2023).
\bibitem{Werner} V Werner et al., Phys. Lett. {\bf {B 550}}, 140 (2002).
\bibitem{Simpson} G. S. Simpson et al., Phys. Rev. {\bf{C 74}}, 064308 (2006).
\bibitem{Nomura} K. Nomura, Phys. Rev. {\bf{C 94}}, 044314 (2016).
\bibitem{Ring} P. Ring and P. Schuck, The nuclear many$-$body problem, Springer$-$Verlag, Berlin (1980).
\bibitem{Crider} B.P. Crider, Physics Letters {\bf{B 763}}, 108 (2016).
\bibitem{GMAJJPGlife} G. Saxena, M. Aggarwal, D. Singh, A. Jain, P. K. Sharma and H. L. Yadav, J. of Phys. G: Nucl. Part. Phys. {\bf {50}}, 015102 (2023).
\bibitem{Hua} H. Hua et. al., Phys. Rev. {\bf{C 69}}, 014317 (2004).
\bibitem{Sarriguren} P. Sarriguren et. al., Phys. Rev. {\bf{C 89}}, 034311 (2014).
\bibitem{Abusara} H. Abusara et al., Phys. Rev {\bf{C 95}}, 054302 (2017).
\bibitem{MAPRB693} M. Aggarwal, Phys. Rev. {\bf{B 693}}, 489 (2010).
\bibitem{MAIJMP28} M. Aggarwal et al., Int. Journ. of Mod Phys E, Vol. {\bf {28}}, No. 11, 1950099 (2019).
\bibitem{MollerBengtsson} P. Möller, R. Bengtsson, B. G. Carlsson, P. Olivius, and T.Ichikawa, Phys. Rev. Lett. {\bf {97}}, 162502 (2006).
\bibitem{Xiang} J. Xiang, J. M. Yao, Y. Fu, Z. H. Wang, Z. P. Li and W. H. Long, Phys. Rev. {\bf {C 93}}, 054324 (2016).
\bibitem{Cejnar} P. Cejnar, J. Jolie, and R. F. Casten, Rev. Mod. Phys. {\bf{82}}, 2155 (2010).
\bibitem{Sumikama} T. Sumikama et. al., Phys. Rev. Lett. {\bf{106}}, 202501 (2011).
\bibitem{Goodin} C. Goodin et al., Nucl. Phys. {\bf{A 787}}, 231 (2007).
\bibitem{Urban} W. Urban et al., Nucl. Phys. {\bf{A 689}}, 605 (2001).
\bibitem{Campbell} P. Campbell et al., Phys. Rev. Lett. {\bf{89}}, 082501 (2002).
\bibitem{Charlwood} F. C. Charlwood et al., Phys. Lett.{\bf{ B 674}}, 23 (2009).
\bibitem{Heyde} K. Heyde and J. L. Wood, Rev. Mod. Phys. {\bf{83}}, 1655 (2011).
\bibitem{Butler} P. A. Butler and W. Nazarewicz, Rev. Mod. Phys. {\bf{68}}, 349 (1996).
\bibitem{Sorlin} O. Sorlin and M. G. Porquet,Prog. Part. Nucl. Phys. {\bf{61}}, 602 (2008).
\bibitem{Bender} M. Bender, P.H. Heenen, and P.G. Reinhard, Rev. Mod. Phys. {\bf{75}}, 121 (2003).
\bibitem{Marshalek} E. R. Marshalek, Nucl. Phys. {\bf{A 331}}, 429 (1979).
\bibitem{Odegard} S. W. Ødegård et al., Phys. Rev. Lett. {\bf{86}} , 5866 (2001).
\bibitem{Frauendorf} S. Frauendorf and J. Meng, Nucl. Phys. {\bf{A 617}}, 131 (1997).
\bibitem{Grodner} E. Grodner et al., Phys. Rev. Lett. {\bf{97}}, 172501 (2006).
\bibitem{Meng} J. Meng, J. Peng, S. Q. Zhang, and S. G. Zhou, Phys. Rev. {\bf{C 73}}, 037303 (2006).
\bibitem{Ayangeakaa} A. D. Ayangeakaa et al., Phys. Rev. Lett. {\bf{110}}, 172504 (2013).
\bibitem{Stachel} J. Stachel, N. Kaffrell, E. Grosse, H. Emling, H. Folger, R. Kulessa, and D. Schwalm, Nucl. Phys. {\bf{A 383}}, 429 (1982).
\bibitem{Zamfir} N. Zamfir and R. Casten, Phys. Lett. {\bf{B 260}}, 265 (1991).
\bibitem{Sheikh} J. A. Sheikh, Y. Sun, and R. Palit, Phys. Lett. {\bf{B 507}}, 115 (2001).
\bibitem{Yao} J. M. Yao, K. Hagino, Z. P. Li, J. Meng, and P. Ring, Phys. Rev. {\bf{C 89}}, 054306 (2014).
\bibitem{Fu} Y. Fu, H. Mei, J. Xiang, Z. P. Li, J. M. Yao, and J. Meng, Phys. Rev. {\bf{C 87}}, 054305 (2013).	
\bibitem{Lu} B. N. Lu, E.G. Zhao, and S.G. Zhou, Phys. Rev. {\bf{C 85}}, 011301(R) (2012).
\bibitem{Dauphas} N. Dauphas et al., Earth \& Planetary Science Letters, Volume {\bf{226}}, Issues 3$-$4, 465 (2004).
\bibitem{Hukkanen} M. Hukkanen et al. Phys. Rev. {\bf{C 108}}, 064315 (2023).
\bibitem{Karmakar} A. Karmakar et al., Arxiv {\bf{2306.07670}} (2022).
\bibitem{MAPLB} M. Aggarwal, Phys. Lett.{\bf{B 693}}, 489 (2010).
\bibitem{MAPRC90} M. Aggarwal, Phys. Rev. {\bf{C 90}}, 064322 (2014).
\bibitem{saxenaPLB2019} G. Saxena, M. Kumawat, M. Kaushik, S.K. Jain, Mamta Aggarwal, Phys. Lett. B \textbf{788}, 1 (2019).
\bibitem{saxena2017} G. Saxena, M. Kumawat, M. Kaushik, U. K. Singh, S. K. Jain, S. Somorendro Singh, M. Aggarwal, Int. J. Mod. Phys. E {\bf{26}}, 1750072 (2017).
\bibitem{saxenaplb2017} G. Saxena, M. Kumawat, M. Kaushik, S.K. Jain, M. Aggarwal,Phys. Lett. B \textbf{775}, 126 (2017).
\bibitem{lala} G. A. Lalazissis et. al., Phys. Lett. {\bf{B671}}, 36$-$41, (2009).
\bibitem{bhuyan2015} M. Bhuyan, Phys. Rev. {\bf{C 92}}, 034323 (2015).
\bibitem{shi2018} Z. Shi et. al, Phys. Rev. {\bf{C 97}}, 034329 (2018).
\bibitem{niksic08} T. Nikšić, D. Vretenar, and P. Ring, {\bf{C 78}}, 034318 (2008).
\bibitem{Brack} M. Brack, J. Damgaard, A. S. Jensen, H. C. Pauli, V. M. Strutinsky, and C. Y. Wong, Rev. Mod. Phys. {\bf{44}}, 320 (1972).
\bibitem{GsN} G. Shanmugam, P. R. Subramanian, M. Rajasekaran, and V. Devanathan, Nuclear Interactions, {\bf{Vol. 72}}, 433 (Springer, Berlin,1979).
\bibitem{GsL} G. Shanmugam, P. R. Subramanian, M. Rajasekaran, and V. Devanathan, Lecture Notes in Physics, {\bf{Vol. 72}}, 433 (Springer, Berlin,1979).
\bibitem{Eis} J. M. Eisenberg and W. Greiner, Vol. 3 of Nuclear Theory (North Holland, New York, 1976).
\bibitem{Seg} P. A. Seeger and W.M. Howard, Nucl. Phys. {\bf{A 238}}, 491 (1975).	
\bibitem{Pm} P. Moller, J. R. Nix, W. D. Myers and W. J. Swiatecki, At. Data Nucl. Data Tables, {\bf{59}}, 185 (1995).
\bibitem{Lalazissis09} G. A. Lalazissis, S. Karatzikos, R. Fossion, D. Pena Arteaga, A. V. Afanasjev and P. Ring, \textit{Phys. Lett. B} \textbf{671}, 36 (2009).
\bibitem{Lalazissis05} G. A. Lalazissis, T. Niksic, D. Vretenar, and P. Ring,   \textit{Phys. Rev. C} \textbf{71}, 024312 (2005).
\bibitem{agbemava2014} S. E. Agbemava, A. V. Afanasjev, D. Ray, and P. Ring, \textit{Phys. Rev. C} \textbf{89}, 054320 (2014).
\bibitem{Boguta77} J. Boguta and A. R. Bodmer, \textit{Nucl. Phys. A} \textbf{292}, 413 (1977) .
\bibitem{Boguta83} J. Boguta and H. Stoecker, \textit{Phys. Lett. B} \textbf{120}, 289 (1983).
\bibitem{Furnstahl97} R. J. Furnstahl, B. D. Serot, and H. B. Tang,  \textit{Nucl. Phys. A} \textbf{615}, 441 (1997) .
\bibitem{Geng2003}L.S. Geng, H. Toki, S. Sugimoto and J. Meng, \textit{Prog. Theor. Phys.} \textbf{110}, 921 (2003) .
\bibitem{Gambhir1989}Y. K. Gambhir, P. Ring and A. Thimet, \textit{Annals Phys.} \textbf{198}, 132 (1990) .
\bibitem{Ring1996} P. Ring, \textit{Prog. Part. Nucl. Phys.} \textbf{37}, 193 (1996).
\bibitem{Yadav2004} H. L. Yadav, M. Kaushik, and H. Toki, \textit{Int. J. Mod. Phys. E} \textbf{13}, 647 (2004) .
\bibitem{Dobaczewski1983}J. Dobaczewski, H. Flocard and J. Treiner, \textit{Nuclear Physics A} \textbf{422} 103 (1984).
\bibitem{Bertsch1991}G. F. Bertsch and H. Esbensen, \textit{Annals Phys.} \textbf{209}, 327 (1991).
\bibitem{ringcpc1997} P. Ring, Y. K. Gambhir, and G. A. Lalazissis, Comput. Phys. Commun. {\bf{A 105}}, 77 (1997).
\bibitem{Singh2013} D. Singh, G. Saxena, M. Kaushik, H. L. Yadav and H. Toki, \textit{Int. J. Mod. Phys. E} \textbf{22}, 1392001 (2013).
\bibitem{TMR} Y. Tian, Z. Y. Ma, and P. Ring,  \textit{Phys. Lett. B} \textbf{676}, 44 (2009).
\bibitem{berger1991} J. F. Berger, M. Girod, and D. Gogny, \textit{Comp. Phys. Comm.} \textbf{63}, 365 (1991).
\bibitem{niksiccpc2014} T. Nikšić, N. Paar, D. Vretenar and P. Ring, Comput. Phys. Commun. {\bf{185}}, 1808 (2014).
\bibitem{Wood} J. L. Wood, K. Heyde, W. Nazarewicz, M. Huyse, and P. van Duppen, Phys. Rep. {\bf{A 215}}, 101 (1992).
\bibitem{Julin} R. Julin, K. Helariutta, and M. Muikku, J. Phys. {\bf{G 27}}, R109 (2001).
\bibitem{Wapstra} F.G. Kondev, M. Wang, W.J. Huang, S. Naimi and G. Audi, , Chinese J. of Phys. {\bf C 45}, 030001 (2021).
\bibitem{ws4} N. Wang, M. Liu, X. Wu and J. Meng 2014, \textit{Phys. Lett. B} \textbf{734} 215 (2014).
\bibitem{ktuy2005} H. Koura, T. Tachibana, M. Uno, M. Yamada, Progress of Theoretical Physics, Volume {\bf{113}}, Issue 2, 305$-$325 (2005).
\bibitem{Manpald} Mamta Aggarwal, Nucl. Phys. {\bf{A 1032}}, 122619, (2023).
\bibitem{Erhard} M. Erhard et. al., Phys. Rev {\bf{C 81}}, 034319 (2010).
\bibitem{Raman} S. Raman, C. W. Nestor, JR., and P. Tikkanen, T Atomic Data and Nuclear Data Tables {\bf{78}}, 1 (2001).
\bibitem{Dhiman} S. Dhiman et al., Nucl. Phys. {\bf{A 1014}}, 122254 (2021).
\bibitem{Rodriguez} R. Rodríguez$-$Guzmán, P. Sarriguren a, L.M. Robledo b and S. Perez$-$Martin, Phys. Lett. {\bf{B 691}}, 4, 202 (2010).
\bibitem {GMAJOGMG} G. Saxena et al, J. Phys. G: Nucl. Part. Phys. {\bf{48}}, 125102 (2021)
\bibitem{Nishimura2012} S. Nishimura, Prog. Theor. Exp. Phys. Vol. {\bf{2012}}, 03C006 (2012).
\bibitem{Nishimura2011} S. Nishimura et.al., Phys. Rev. Lett. {\bf{106}}, 052502 (2011).
\bibitem{sobhani2022}  H. Sobhani and H. Khalafi, Chin. J. of Phys. vol. {\bf{85}}, 475 (2023) (https://doi.org/10.1016/j.cjph.2022.10.011).

\end{thebibliography}
\end{document}